\documentclass[aps,prd,showpacs,preprintnumbers,amsmath,amssymb,superscriptaddress,preprintstyle,preprint,nofootinbib]{revtex4-1}
\usepackage{graphicx}
\usepackage{dcolumn}
\usepackage{bm}
\usepackage{color}
\usepackage{soul,xcolor}
\usepackage[utf8]{inputenc}
\usepackage{epsfig}
\usepackage{epsf}
\usepackage{dcolumn}
\def\be{\begin{equation}}
\def\ee{\end{equation}}
\def\bea{\begin{eqnarray}}
\def\eea{\end{eqnarray}}
\def\gsim{\ \rlap{\raise 2pt\hbox{$>$}}{\lower 2pt \hbox{$\sim$}}\ }
\def\lsim{\ \rlap{\raise 2pt\hbox{$<$}}{\lower 2pt \hbox{$\sim$}}\ }
\def\dslash{\kern-4pt \not{\hbox{\kern-2pt $\partial$}}}
\def\pslash{\not{\hbox{\kern-2pt p}}}

\def\pmue{{{P_{\mu e}} }}

\newcommand{\nue}{\nu_e}
\newcommand{\numu}{\nu_\mu}
\newcommand{\nutau}{\nu_\tau}

\newcommand{\epssme}{\varepsilon^s_{\mu e}}
\newcommand{\epssmm}{\varepsilon^s_{\mu\mu}}
\newcommand{\epssmt}{\varepsilon^s_{\mu\tau}}

\newcommand{\epsdee}{\varepsilon^d_{ee}}
\newcommand{\epsdem}{\varepsilon^d_{e\mu}}

\newcommand{\epsdme}{\varepsilon^d_{\mu e}}
\newcommand{\epsdmm}{\varepsilon^d_{\mu\mu}}

\newcommand{\epsdte}{\varepsilon^d_{\tau e}}
\newcommand{\epsdtm}{\varepsilon^d_{\tau\mu}}

\newcommand{\epsmee}{\varepsilon^m_{ee}}

\newcommand{\epsmme}{\varepsilon^m_{\mu e}}
\newcommand{\epsmmm}{\varepsilon^m_{\mu\mu}}

\newcommand{\epsmte}{\varepsilon^m_{\tau e}}
\newcommand{\epsmtm}{\varepsilon^m_{\tau\mu}}
\newcommand{\epsmtt}{\varepsilon^m_{\tau\tau}}

\newcommand{\dcp}{\delta}
\newcommand{\nova}{NO$\nu$A }
\newcommand{\dm}[1]{\Delta m^2_{#1}}

\setstcolor{blue}

\begin{document}
\DeclareGraphicsExtensions{.eps,.ps}


\title{A combined study of source, detector and matter non-standard neutrino
interactions at DUNE}



\author{Mattias~Blennow}
\email[Email Address: ]{emb@kth.se}
\affiliation{
Department of Theoretical Physics,
School of Engineering Sciences, KTH Royal Institute of Technology,
AlbaNova University Center, 106 91 Stockholm, Sweden
}
 
\author{Sandhya~Choubey}
\email[Email Address: ]{sandhya@hri.res.in}
\affiliation{
Department of Theoretical Physics,
School of Engineering Sciences, KTH Royal Institute of Technology,
AlbaNova University Center, 106 91 Stockholm, Sweden
}
\affiliation{
Harish-Chandra Research Institute,
Chhatnag Road, Jhunsi, Allahabad 211 019, India
}
 
\author{Tommy~Ohlsson}
\email[Email Address: ]{tohlsson@kth.se}
\affiliation{
Department of Theoretical Physics,
School of Engineering Sciences, KTH Royal Institute of Technology,
AlbaNova University Center, 106 91 Stockholm, Sweden
}

\author{Dipyaman~Pramanik}
\email[Email Address: ]{dipyamanpramanik@hri.res.in}
\affiliation{
Harish-Chandra Research Institute,
Chhatnag Road, Jhunsi, Allahabad 211 019, India
}
 
\author{Sushant~K.~Raut}
\email[Email Address: ]{raut@kth.se}
\affiliation{
Department of Theoretical Physics,
School of Engineering Sciences, KTH Royal Institute of Technology,
AlbaNova University Center, 106 91 Stockholm, Sweden
}

\begin{abstract}
We simultaneously investigate source, detector and matter non-standard neutrino interactions at
the proposed DUNE experiment. Our analysis is performed using a Markov Chain 
Monte Carlo exploring the full parameter space. We find that the sensitivity of 
DUNE to the standard oscillation parameters is worsened due to the presence of 
non-standard neutrino interactions. In particular, there are degenerate solutions in the 
leptonic mixing angle $\theta_{23}$ and the Dirac CP-violating phase $\dcp$. 
We also compute the expected sensitivities at DUNE to the 
non-standard interaction parameters. We find that the sensitivities to the matter 
non-standard interaction parameters are substantially stronger than the current 
bounds (up to a factor of about 15). Furthermore, we discuss correlations between the source/detector and matter 
non-standard interaction parameters and find a degenerate solution in $\theta_{23}$. 
Finally, we explore the effect of statistics 
on our results. 
\end{abstract}
\maketitle

\section{Introduction}

Despite its unprecedented success, it is now well established that the Standard
Model of elementary particles (SM) needs to be extended. Two of the main
observational evidences that cannot be described in the context of the SM are
the existence of neutrino masses and mixing and the presence of dark matter in
the universe. A plethora of models proposing an extension of SM has been put
forth that describe non-zero neutrino mass and mixing and/or dark matter. Some
of these new physics models could, in principle, lead to corrections to the
effective neutrino interaction vertices through additional (higher-order) terms.
All such new charged-current (CC) interactions could therefore lead to
modifications in the production and detection of neutrinos, while all
neutral-current (NC) interactions  could modify the neutrino-matter forward
scattering cross-section and hence affect neutrino oscillations in matter.
Without going into details of the ultraviolet-complete models, these
so-called non-standard 
interactions (NSIs) involving neutrinos can be parametrised in terms of 
effective four-fermion operators (see e.g.~Ref.~\cite{Ohlsson:2012kf} and
references therein).
Hence, the NSI parameters are of two kinds: the 
source/detector NSIs (for CC) and the matter NSIs (for NC), respectively. 
They have been studied extensively both in the context of existing constraints
on the effective NSI parameters as well as in the context of expected
constraints coming from future neutrino oscillation experiments. If present,
NSIs could also fake expected event spectra due to standard neutrino
oscillations, but with a different set of parameter values. This could therefore
lead to new degeneracies in the oscillation parameter space. An important
challenge for future experiments is thus to find ways to break these
degeneracies to obtain maximum sensitivity to oscillation parameters. 

Discovery of CP violation in the lepton sector is amongst the most important
goals for future neutrino oscillation experiments. The Deep Underground Neutrino
Experiment (DUNE)
\cite{Acciarri:2016crz,Acciarri:2015uup,Strait:2016mof,Acciarri:2016ooe} is
being proposed as a discovery set-up for CP violation. It is expected to have
all ingredients needed for precision search of the Dirac CP-violating phase
$\delta$, a powerful beam to be built at Fermilab, a large high-end liquid argon
detector at Sanford Underground Research Facility (SURF), near detector to
control detector systematics, and a long baseline of 1300 km from Fermilab in
Illinois, USA to SURF in South Dakota, USA. The physics reach of DUNE
in the presence of standard oscillations has been studied extensively by
the DUNE collaboration \cite{Acciarri:2015uup} and others. Any sizable new
physics effect is expected to modify the event spectrum at the DUNE detector,
and hence, its physics reach. Effect of matter NSIs on DUNE has been previously studied 
in
Refs.~\cite{Masud:2015xva,deGouvea:2015ndi,Coloma:2015kiu,Masud:2016bvp,
Masud:2016gcl}. 

It has been established that the presence of matter NSIs in general reduces the
sensitivity of DUNE to standard oscillation parameters. The main reason behind
this reduction is the interplay between oscillations due to standard and
non-standard parameters that give rise to new kinds of degeneracies for long
baseline experiments \cite{deGouvea:2015ndi,Coloma:2015kiu}. It has been shown
\cite{Coloma:2016gei} that for sufficiently large values of the NSI parameters
one could expect a degeneracy between the sign of $\Delta m_{31}^2$ and
$\delta$, affecting the sensitivity of DUNE to the neutrino mass ordering
\cite{Coloma:2016gei,Masud:2016gcl}. For $\theta_{23}$ and CP measurements,
studies have revealed that there are two other degeneracies. The first kind is
due to an interplay between the oscillation parameter $\theta_{23}$ and the NSI
parameters. 
This leads to a reduction of the DUNE sensitivity to $\theta_{23}$ 
and even fake so-called octant solutions~\cite{deGouvea:2015ndi,Coloma:2015kiu}.
The second kind is due to an interplay between $\delta$ and the NSI parameters,
opening up the possibility of a reduced expected sensitivity for this parameter
at DUNE. Since the NSI paradigm brings in a large number of parameters, the
statistical analysis of the projected data at DUNE becomes cumbersome and
challenging. The analysis with a full matter NSI parameter scan was performed in
Refs.~\cite{deGouvea:2015ndi,Coloma:2015kiu} for a three years running of the
experiment in the neutrino mode and three years in the antineutrino mode. To the best of
our knowledge, the impact of source and detector NSIs at DUNE has not been
studied before.

 Any theory, which gives rise to the matter NSIs, would almost always also give
rise to source and detector NSIs, and hence, it is imperative to consider them
together in a complete analysis. The neutrino oscillation probabilities in
presence of both source/detector and matter NSIs have been calculated
before \cite{Kopp:2007ne} and are seen to depend on these parameters in a
correlated way. It is therefore pertinent to ask if these correlation could
alter in any way the expected sensitivity of DUNE. 

In this paper, we perform a
complete analysis of the expected sensitivity of DUNE, allowing for both source/detector 
and matter NSIs. 
We study the combined effect of source/detector and matter NSIs and look at possible correlations between
them at the level of oscillation probabilities. We point out the importance of
the event spectrum in disentangling standard oscillations from oscillations
driven by NSI parameters. We next calculate the expected sensitivity of DUNE
for 
standard and NSI parameters from a full scan of
the NSI parameter space, including all relevant source/detector and
matter NSIs. Finally, we explore the effect of runtime on the
precision measurement at DUNE.

\section{Neutrino oscillations with non-standard interactions}
\label{sec:neutosc}

The presence of flavour off-diagonal operators beyond the SM 
is manifest in the phenomenon of neutrino oscillations. 
In the standard picture of neutrino oscillations, a neutrino produced 
at a source in association with a charged lepton $\ell_\alpha$ is simply
\begin{equation}
 | \nu_\alpha^s \rangle  =  | \nu_\alpha \rangle~,
\end{equation}
i.e.~the weak-interaction eigenstate with isospin $T^3=+1/2$. 
Similarly, a neutrino that produces a charged lepton $\ell_\beta$ 
at a detector is 
\begin{equation}
 \langle \nu_\beta^d |  =  \langle \nu_\beta | ~,
\end{equation}
which is also the weak-interaction eigenstate. Between the source and 
the detector, the propagation of neutrinos with energy $E$ is governed by the
time-evolution 
equation 
\begin{equation}
 \textrm{i} \frac{\textrm{d}}{\textrm{d} t} \left[ \begin{array}{c}
                                \nue \\ \numu \\ \nutau
                               \end{array} \right] = \frac{1}{2E} \left\{
U^\dagger \left[ \begin{array}{ccc}
                                                                                
           0 & 0 & 0 \\
                                                                                
           0 & \dm{21} & 0 \\
                                                                                
           0 & 0 & \dm{31}
                                                                                
          \end{array} \right] U + \left[ \begin{array}{ccc}
                                                                                
                                          A & 0 & 0 \\
                                                                                
                                          0 & 0 & 0 \\
                                                                                
                                          0 & 0 & 0
                                                                                
                                         \end{array} \right] \right\} \left[
\begin{array}{c}
										
								      \nue \\
\numu \\ \nutau

\end{array} \right] ~.
\end{equation}
Here, $U$ is the leptonic mixing matrix that is parametrized in terms of 
three mixing angles $\theta_{12}$, $\theta_{13}$ and $\theta_{23}$ and one 
Dirac CP-violating phase $\dcp$. The evolution of neutrino states also depends
on 
the two independent mass-squared differences $\dm{ij}=m_i^2-m_j^2$. When 
neutrinos propagate through the earth, the coherent forward scattering of
$\nue$ 
off electrons results in the matter potential $A = 2\sqrt{2}G_F n_e E$, where 
$n_e$ is the number density of electrons. Thus, standard neutrino oscillation 
probabilities depend on six oscillation parameters, and are modified by 
matter effects. 

Beyond the SM, it is possible to have CC-like 
operators that affect the interactions of neutrinos with charged leptons. 
If these operators are not diagonal in flavour basis, then the 
production and the detection of neutrinos are affected. The neutrino state 
produced at the source in association with the charged lepton $\ell_\alpha$ 
then also has components of the other flavours 
\begin{equation}
 | \nu_\alpha^s \rangle  =  | \nu_\alpha \rangle  +  \sum_{\gamma=e,\mu,\tau}
\varepsilon^s_{\alpha\gamma}  | \nu_\gamma \rangle  ~,
\end{equation}
and similarly at the detector, 
\begin{equation}
 \langle \nu_\beta^d |  =  \langle \nu_\beta |  +  \sum_{\gamma=e,\mu,\tau}
\varepsilon^d_{\gamma\beta}  \langle \nu_\gamma |  ~.
\end{equation}
The matrices $\varepsilon^s = (\varepsilon^s_{\alpha\gamma})$ and $\varepsilon^d = (\varepsilon^d_{\gamma\beta})$ that represent the 
source and the detector NSIs, repectively, 
are in general complex matrices with 18 real parameters each. These are 
the nine amplitudes $|\varepsilon^{s/d}_{\alpha\beta}|$ and nine phases 
$\varphi^{s/d}_{\alpha\beta}$. Note that the definitions of $\varepsilon^s_{\alpha\gamma}$ 
and $\varepsilon^d_{\gamma\beta}$ follow the convention used in 
Ref.~\cite{Ohlsson:2012kf}.

The NC-like operators affect the propagation of 
neutrinos through matter, inducing more terms similar to the matter 
potential. The modified time-evolution equation is 
\begin{equation}
 \textrm{i} \frac{\textrm{d}}{\textrm{d} t} \left[ \begin{array}{c}
                                \nue \\ \numu \\ \nutau
                               \end{array} \right] = \frac{1}{2E} \left\{
U^\dagger \left[ \begin{array}{ccc}
                                        0 & 0 & 0 \\
                                        0 & \dm{21} & 0 \\
                                        0 & 0 & \dm{31}
                                       \end{array} \right] U + A \left[
\begin{array}{ccc}
                                                                        
1+\varepsilon^m_{ee} & \varepsilon^m_{e\mu} & \varepsilon^m_{e\tau} \\
                                                                        
\varepsilon^m_{\mu e} & \varepsilon^m_{\mu\mu} & \varepsilon^m_{\mu\tau} \\
                                                                        
\varepsilon^m_{\tau e} & \varepsilon^m_{\tau\mu} & \varepsilon^m_{\tau\tau}
									
\end{array} \right] \right\}  \left[ \begin{array}{c}
										
								      \nue \\
\numu \\ \nutau

\end{array} \right]~.
\end{equation}
The entry $1$ in the $e-e$ position of the matter effect matrix stands 
for the standard matter effect, while
the parameters $\varepsilon^m_{\alpha\beta}$ represent the matter NSIs. 
Note that the definitions of $\varepsilon^m_{\alpha\beta}$ also follow 
the convention used in Ref.~\cite{Ohlsson:2012kf}.
Since 
the Hamiltonian has to be Hermitian, we have the relations 
$\varepsilon^m_{\alpha\beta}={\varepsilon^m_{\beta\alpha}}^*$. Thus, there are 
six amplitudes and three phases, i.e.~nine real parameters in the matter NSI 
matrix. Subtracting a constant multiple of the identity matrix does not 
affect the eigenvectors, and hence, oscillation probabilities. Therefore, we
subtract 
the element $\varepsilon^m_{\mu\mu}$ from all the diagonal elements. We define 
${\epsmee}'=\epsmee-\epsmmm$ and ${\epsmtt}'=\epsmtt-\epsmmm$ and treat 
these two new parameters as the physical parameters of the system. 

A comprehensive study of the bounds on NSI parameters has been carried out by 
the authors of Ref.~\cite{Biggio:2009nt}. The 90~\% bounds on the
source/detector NSI parameters{\footnote{Strictly speaking, the 
NSI parameters that affect neutrino oscillations are combinations 
of the NSI parameters that enter the Lagrangian, depending on the 
Lorentz structure of the current involved in the process. In this study, 
we assume for the sake of simplicity that the bounds from Ref.~\cite{Biggio:2009nt} 
apply directly to the oscillation NSI parameters.}}
from their study are as follows
\begin{equation}
 |\varepsilon^{s/d}_{\alpha\beta}| < \left[ \begin{array}{c c c c c}
                                             0.041 & \quad & 0.025 & \quad &
0.041 \\
                                             0.026 & \quad & 0.078 & \quad &
0.013 \\
                                             0.12 & \quad & 0.018 & \quad &
0.13 
                                            \end{array} \right] ~.
\label{eq:bounds_sd}
\end{equation}
For the matter NSI parameters, we follow the discussion in
Ref.~\cite{Choubey:2015xha}. In 
that paper, the authors have used the bounds from Ref.~\cite{Biggio:2009nt}
along with more recent 
results from SK and MINOS~\cite{Mitsuka:2011ty,Ohlsson:2012kf,Adamson:2013ovz}
to obtain the following bounds
\begin{equation}
 |\varepsilon^{m}_{\alpha\beta}| < \left[ \begin{array}{c c c c c}
                                             4.2 & \quad & 0.3 & \quad & 3.0 \\
                                             0.3 & \quad & - & \quad & 0.04 \\
                                             3.0 & \quad & 0.04 & \quad & 0.15 
                                            \end{array} \right] ~.
\label{eq:bounds_m}
\end{equation}
Throughout this article, we will refer to the bounds listed above as the 
`current bounds'.

\begin{figure}
\begin{tabular}{ccc}
\epsfig{file=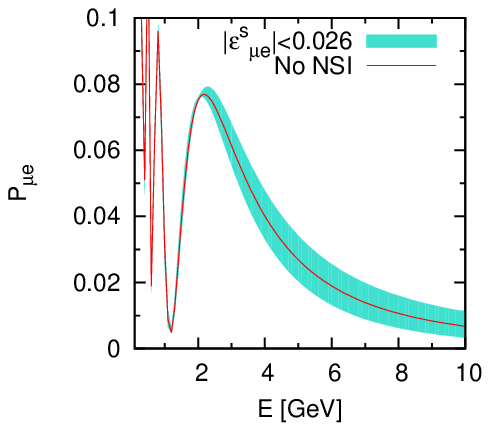, width=0.33\textwidth, bbllx=68, bblly=50, bburx=210,
bbury=176,clip=} &
\epsfig{file=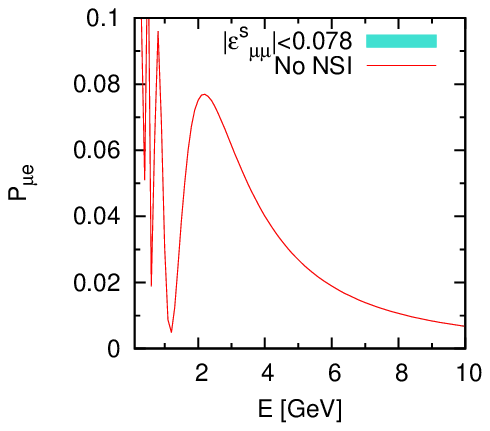, width=0.33\textwidth, bbllx=68, bblly=50, bburx=210,
bbury=176,clip=} & 
\epsfig{file=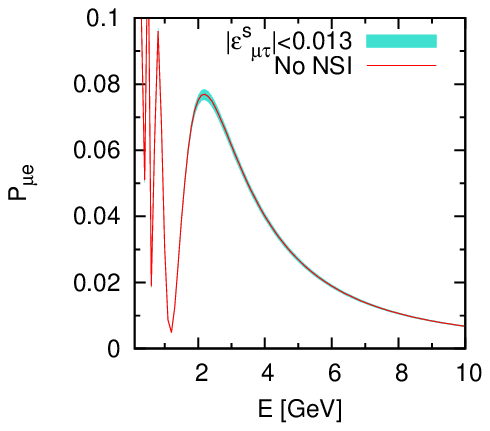, width=0.33\textwidth, bbllx=68, bblly=50, bburx=210,
bbury=176,clip=} \\ 
\epsfig{file=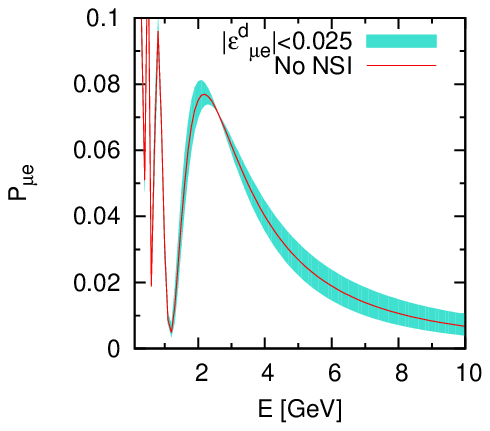, width=0.33\textwidth, bbllx=68, bblly=50, bburx=210,
bbury=176,clip=} &
\epsfig{file=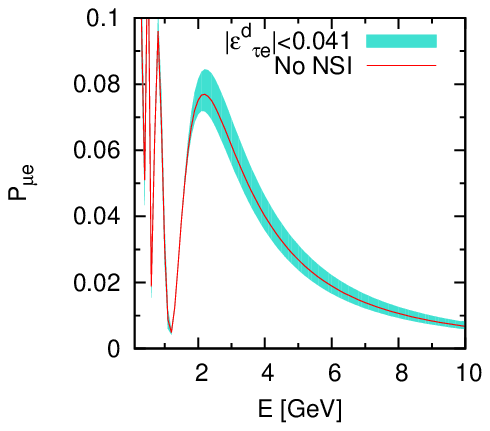, width=0.33\textwidth, bbllx=68, bblly=50, bburx=210,
bbury=176,clip=} & 
 \\ 
\epsfig{file=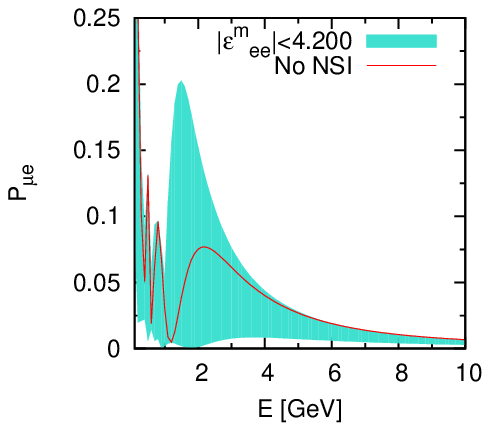, width=0.33\textwidth, bbllx=68, bblly=50, bburx=210,
bbury=176,clip=} &
\epsfig{file=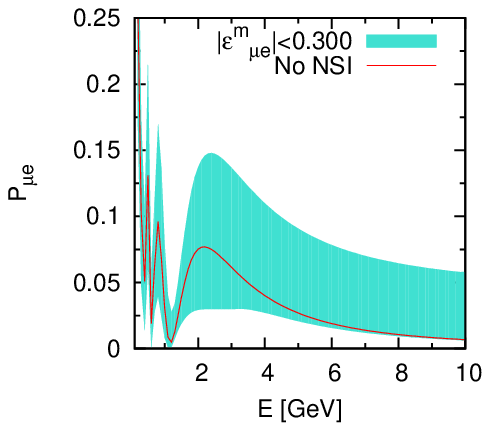, width=0.33\textwidth, bbllx=68, bblly=50, bburx=210,
bbury=176,clip=} & 
\epsfig{file=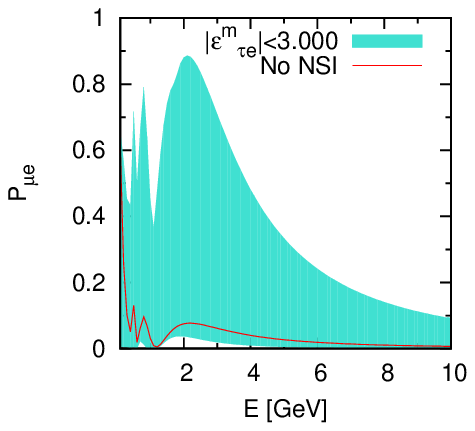, width=0.33\textwidth, bbllx=68, bblly=50, bburx=210,
bbury=176,clip=} \\ 
\epsfig{file=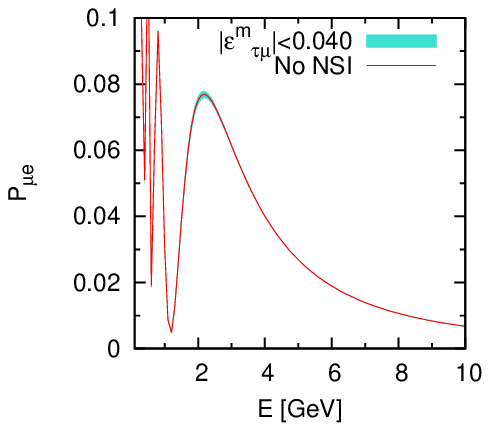, width=0.33\textwidth, bbllx=68, bblly=50, bburx=210,
bbury=176,clip=} &
\epsfig{file=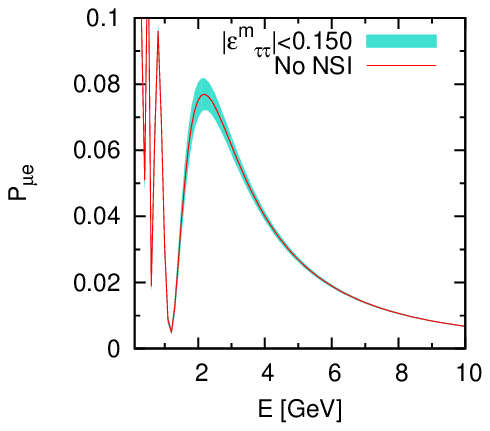, width=0.33\textwidth, bbllx=68, bblly=50, bburx=210,
bbury=176,clip=} & 
\end{tabular}
\caption{\footnotesize Variation in the neutrino oscillation probability $\pmue$
as a function of 
neutrino energy $E$ with some of the NSI parameters varied 
in their allowed range. The central dark curve corresponds to the case of 
no NSIs. The values of the standard oscillation parameters used in generating 
these figures are $\theta_{12}=33.5^\circ$, 
$\theta_{13}=8.48^\circ$, $\theta_{23}=42^\circ$, $\dcp=-90^\circ$, 
$\dm{21}=7.50\times10^{-5} \textrm{ eV}^2$ and $\dm{31}=2.45\times10^{-3}
\textrm{ eV}^2$.}
\label{fig:probs}
\end{figure}

Analytical expressions for the neutrino oscillation probabilities in the 
presence of source/detector and matter NSIs are given in
Ref.~\cite{Kopp:2007ne}. 
The expressions are derived as perturbative expansions in the small 
parameters $\dm{21}/\dm{31}$ and $\sin\theta_{13}$ up to linear order in 
the NSI parameters. In Fig.~\ref{fig:probs}, we show the change in 
the $\numu \to \nue$ oscillation probability $\pmue$ as the NSI parameters 
are varied one at a time within their allowed range at 90~\% C.L. The dark 
curve within the band corresponds to standard oscillations when the value 
of the NSI parameter is zero. Since the existing bounds on matter NSIs are 
weaker, we observe that they affect the probability more, resulting in bands 
that are much wider. Therefore, we expect them to change the event rates at
DUNE 
and affect the measurement of parameters. 

\section{The DUNE experiment}
\label{sec:dune}

DUNE is a proposed long-baseline neutrino oscillation experiment
\cite{Acciarri:2016crz,Acciarri:2015uup,Strait:2016mof,Acciarri:2016ooe}. 
The source of the beam will be at Fermilab, while the liquid argon detector will 
be located at the SURF. The total distance 
travelled by the neutrinos will be around 1300 km. In our simulation, we use
 the neutrino flux corresponding to a 1.2 MW beam with 120 GeV protons. 
The expected flux at DUNE will have beam power between 1.2 MW and 2.3 MW and
proton 
energy between 80 GeV and 120 GeV. Thus, the configuration we are using gives a 
conservative estimate of the net statistics that the experiment will
accumulate. 
Unless stated otherwise, we assume that the experiment will run with five years in 
neutrino mode and five years in antineutrino mode. 

The 40 kiloton liquid argon detector is assumed to have an energy resolution of 15~\% 
for $\nue$ and 20~\% for $\numu$. For NC events, however, the 
reconstructed energy of the neutrino has a wide spread to lower energies, 
because of the production of pions and other hadrons. Therefore, we use 
a smearing matrix to simulate the effect of the reconstruction of these 
events. 

The primary backgrounds to the electron appearance and muon disappearance
signal 
events come from the NC backgrounds and instrinsic $\nue$ contamination in the
flux. 
In addition, there is a wrong-sign component in the flux. The problem of wrong 
sign events is more severe for the antineutrino run, since the neutrino
component 
in the antineutrino flux is larger than the antineutrino component in the
neutrino 
flux. We have taken all of these backgrounds into account in our simulation of
the 
experiment. From an experimental point of view, various cuts are imposed on the 
observed events in order to eliminate as much of the background as possible.
Furthermore, in our 
simulation, the effect of these cuts is to appear as efficiency factors that
reduce the number of events. 
The full specifications of the detector that we have used
can be found in Ref.~\cite{Acciarri:2015uup}. 
Apart from the usual uncertainties in the flux and the cross-section, 
the presence of source/detector NSIs can also affect the calibration of the 
expected number of events. In our analysis, 
we have included systematic errors in the normalization of the flux 
at the 5~\% (20~\%) level for signal (background) events.

\section{Simulation results}
\label{sec:results}

For our simulation, we have made use of the GLoBES package
~\cite{Huber:2004ka,Huber:2007ji} along with 
its auxiliary data files~\cite{messier_xsec,paschos_xsec}. The scanning of the
multi-dimensional parameter 
space was achieved by a Markov Chain Monte Carlo using the MonteCUBES
package~\cite{Blennow:2009pk}. 
As a result, all of our results are to be interpreted in terms of Bayesian
credible regions 
rather than frequentist confidence levels, i.e.~the 90~\% credible region is
the part of 
parameter space that will contain 90~\% of the posterior probability as opposed
to the 
90~\% C.L. contours that contain the points in parameter space where the
simulated 
result would be within the 90~\% least extreme experimental outcomes. We have
written a 
GLoBES-compatible probability engine to handle the full parameter space and 
calculate the oscillation probabilities in the presence of both
source/detector and matter NSIs.

Since we are primarily interested in the $\numu \to \nue$ and $\numu \to \numu$ 
oscillation probabilities, the relevant source NSI parameters are $\epssme$, 
$\epssmm$ and $\epssmt$, while the relevant detector NSI para\-meters are 
$\epsdee$, $\epsdem$, $\epsdme$, $\epsdmm$, $\epsdte$ and $\epsdtm$. The matter 
NSI parameters affect the propagation of neutrinos and are all relevant, 
since intermediate states are summed over. Furthermore, based on the analytical 
expressions and our preliminary simulations, we reduce the set of relevant 
source/detector parameters to $\epssme$, $\epssmm$, $\epssmt$, $\epsdme$ and
$\epsdte$. 
Thus, our final simulations are run over the parameter space spanned by five 
complex source/detector NSI parameters and three complex and two real matter NSI
parameters. 
As for the standard oscillation parameters, we fix the parameters $\dm{21}$ 
and $\theta_{12}$ and vary the others. 

The best-fit values of the standard parameters are $\theta_{12}=33.5^\circ$, 
$\theta_{13}=8.48^\circ$, $\theta_{23}=42^\circ$, $\dcp=-90^\circ$, 
$\dm{21}=7.50\times10^{-5} \textrm{ eV}^2$ and $\dm{31}=2.45\times10^{-3}
\textrm{ eV}^2$ 
which are consistent with the global fits to neutrino oscillation 
data~\cite{Capozzi:2013csa,Forero:2014bxa,Gonzalez-Garcia:2014bfa}. 
These parameters are marginalized 
over their $3\sigma$ ranges allowed by the global fits with the corresponding
priors. For the NSI parameters, we 
use the bounds listed before. The true values of 
these NSI parameters are either set to zero or to a non-zero value equal to 
half of the $1\sigma$ bounds. The true values of all NSI phases are zero, and 
they are free to vary in the entire $[-180^\circ,180^\circ)$ range.

\subsection{Effect on precision measurements at DUNE }

The current generation of long-baseline neutrino oscillation experiments T2K
and 
\nova\ are already collecting data and have provided a hint of the value 
of $\dcp$~\cite{Abe:2015awa,Adamson:2016tbq}. This also gives hints about the
neutrino mass ordering and octant of 
$\theta_{23}$~\cite{Abe:2015awa,Ghosh:2014zea}. If the data collected over the
next few years do not confirm 
these hints, then it may be possible for DUNE to make these measurements. At
any 
rate, we expect that data from DUNE will enable us to measure these unknown 
parameters at a higher confidence level. 

\begin{figure}
\begin{tabular}{cc}
\epsfig{file=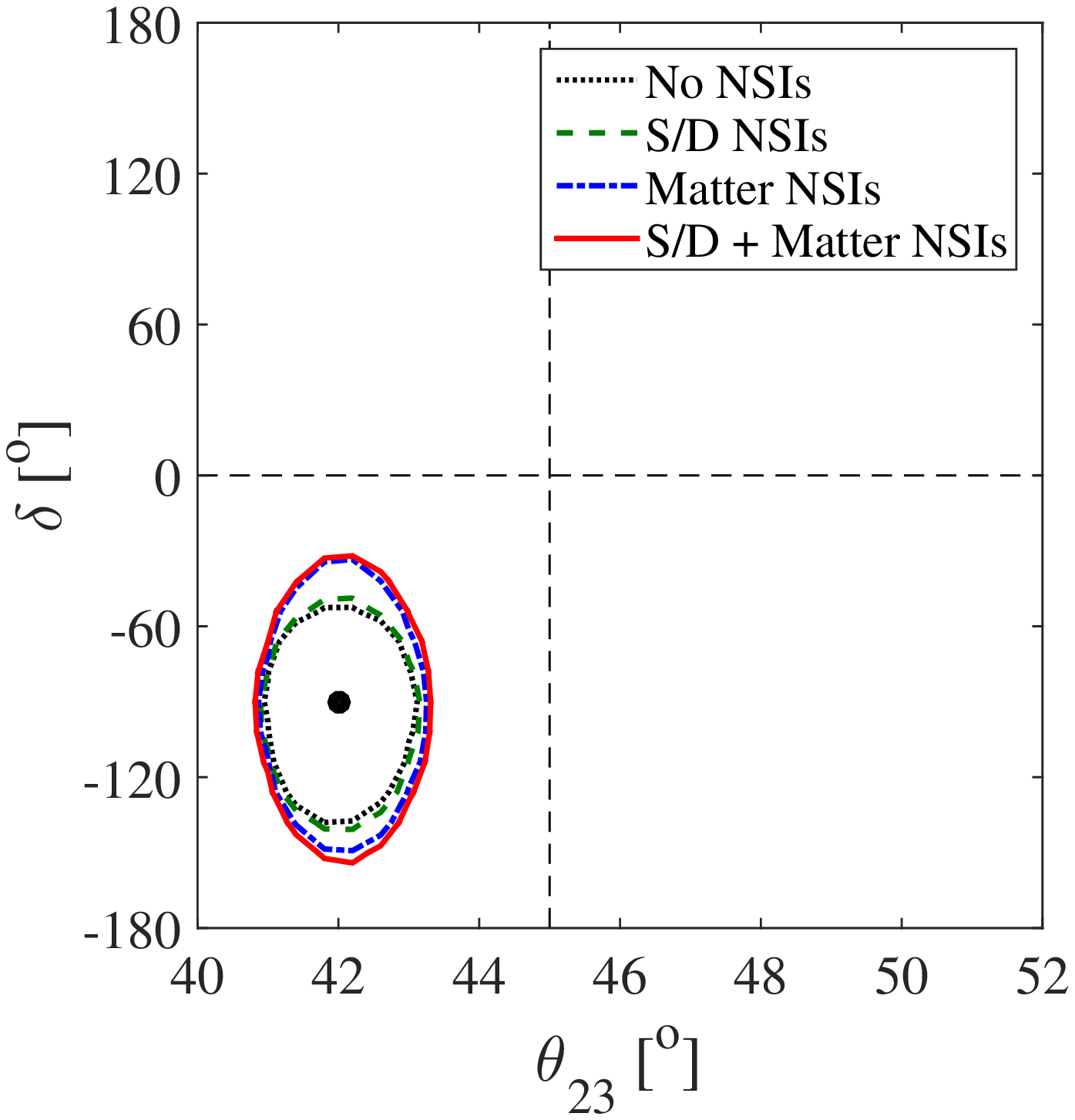, width=0.5\textwidth, bbllx=0, bblly=0, bburx=420,
bbury=420,clip=} &
\epsfig{file=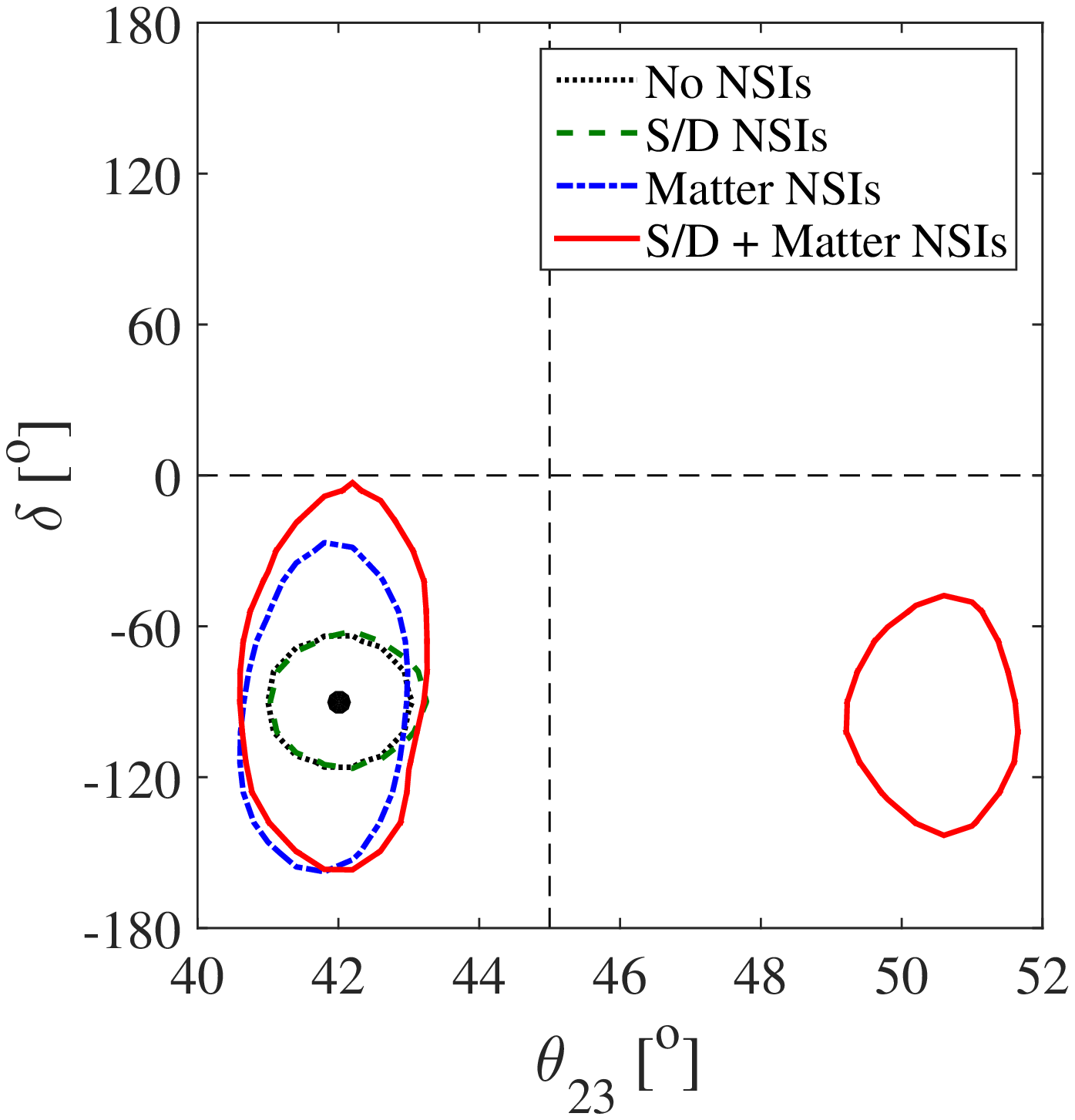, width=0.5\textwidth, bbllx=0, bblly=0,
bburx=420, bbury=420,clip=} 
\end{tabular}
\caption{\footnotesize Sensitivity of DUNE in the $\theta_{23}-\dcp$ plane. The
simulated 
true values of these parameters are $42^\circ$ and $-90^\circ$, respectively. 
The contours enclose the allowed region at 90~\% credible regions obtained by
marginalizing 
over only the standard parameters, standard parameters and 
source/detector NSI parameters, standard parameters and matter NSI parameters, 
and standard parameters and all NSI parameters. In the 
left (right) panel, the true values of the NSI parameters are taken to be 
zero (non-zero).}
\label{fig:effect}
\end{figure}

\begin{figure}
\begin{tabular}{ccc}
\epsfig{file=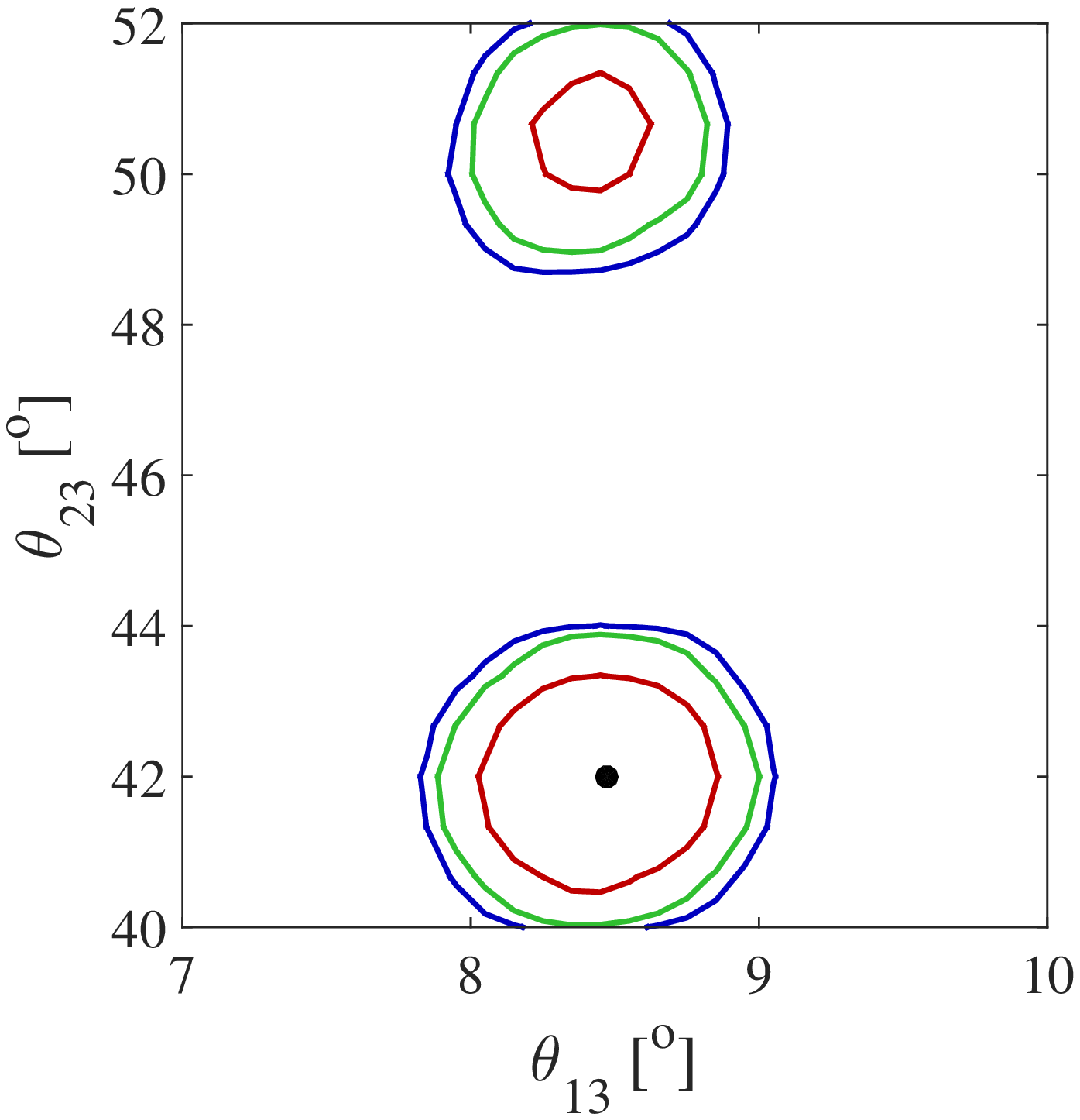, height=0.18\textheight, width=0.33\textwidth,
bbllx=0, bblly=0, bburx=480, bbury=440,clip=} &
&
 \\
\epsfig{file=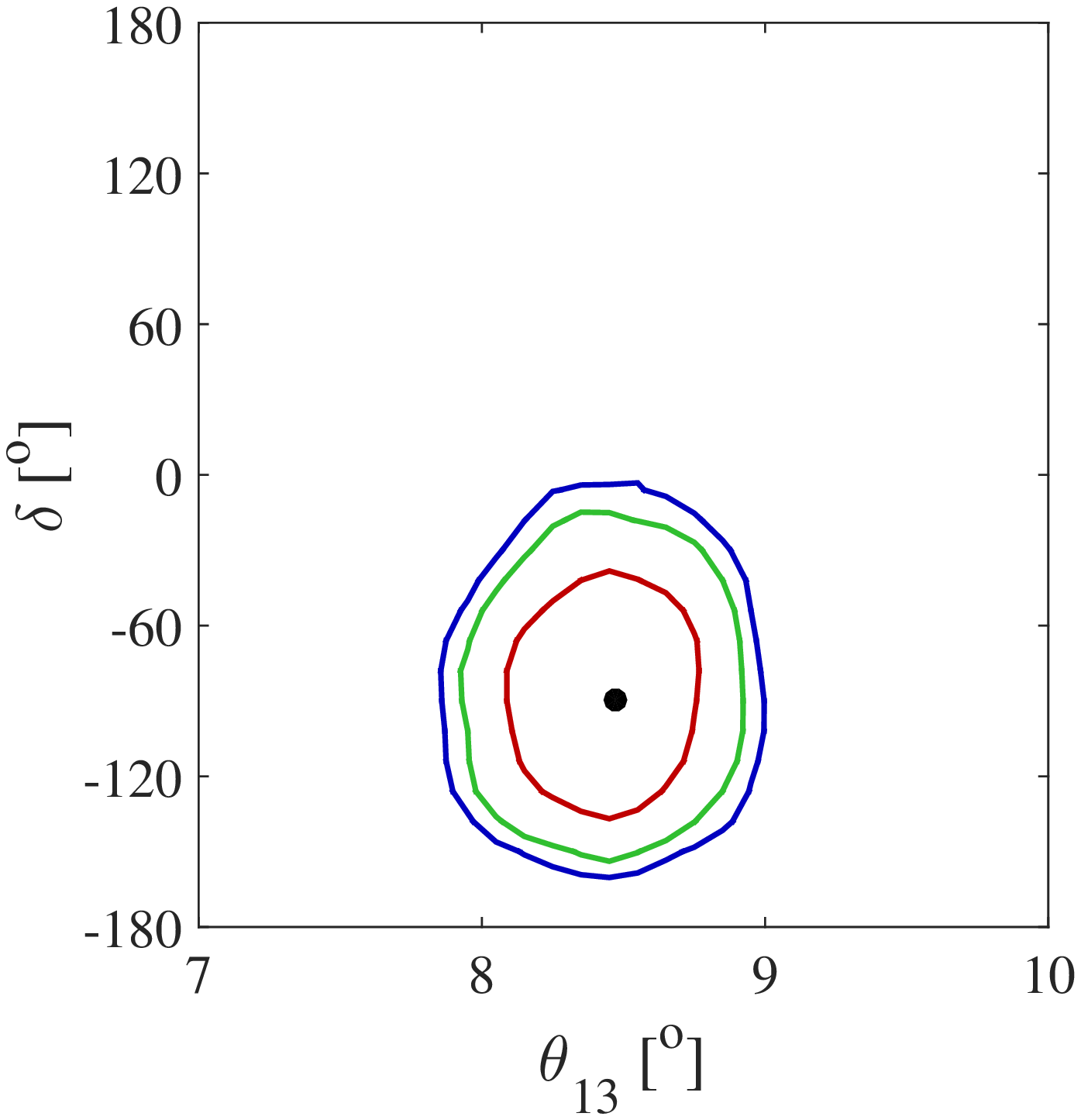, height=0.18\textheight, width=0.33\textwidth,
bbllx=0, bblly=0, bburx=480, bbury=440,clip=} &
\epsfig{file=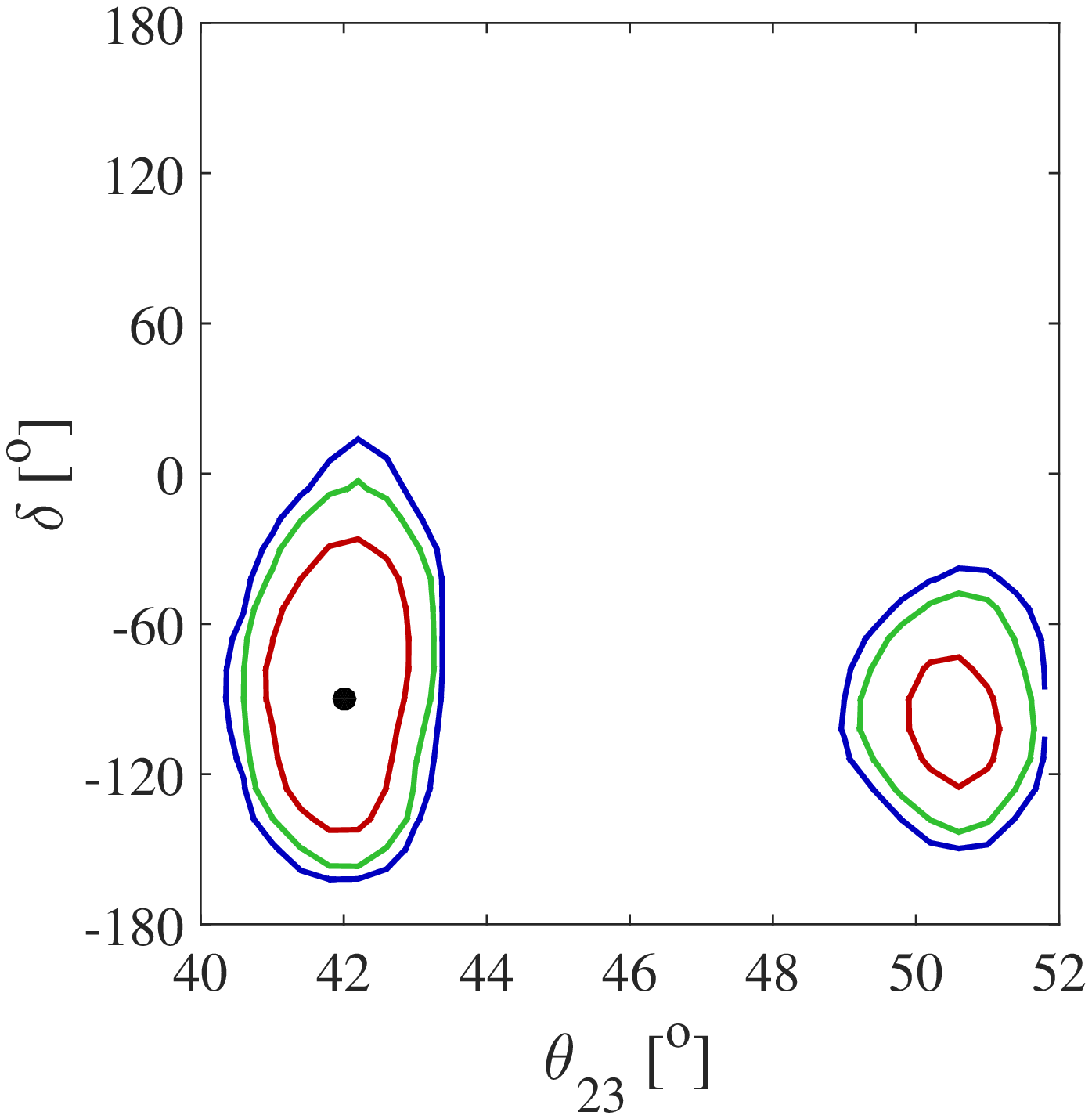, height=0.18\textheight, width=0.33\textwidth,
bbllx=0, bblly=0, bburx=480, bbury=440,clip=} &
 \\
\epsfig{file=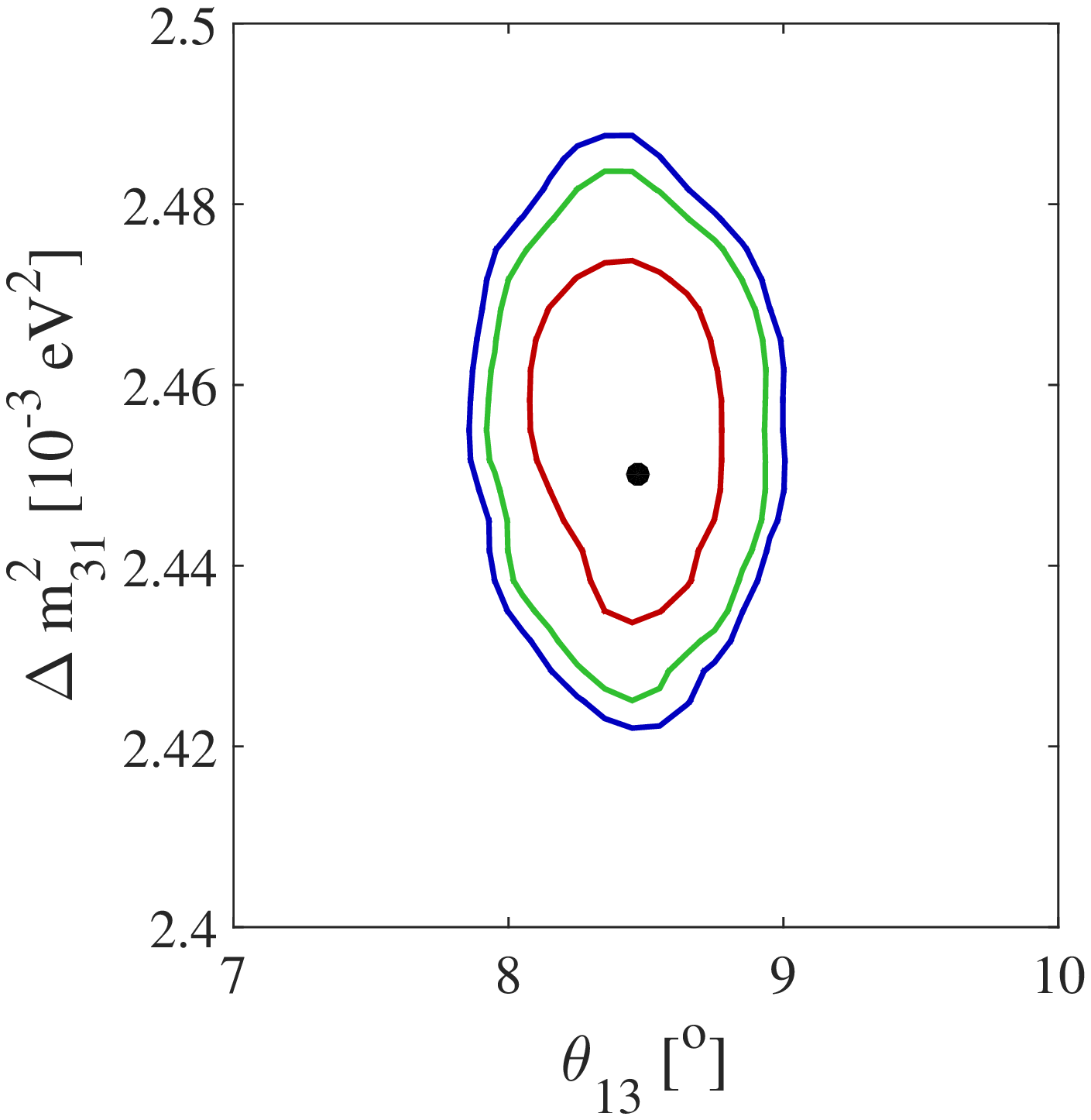, height=0.18\textheight, width=0.33\textwidth,
bbllx=0, bblly=0, bburx=480, bbury=440,clip=} &
\epsfig{file=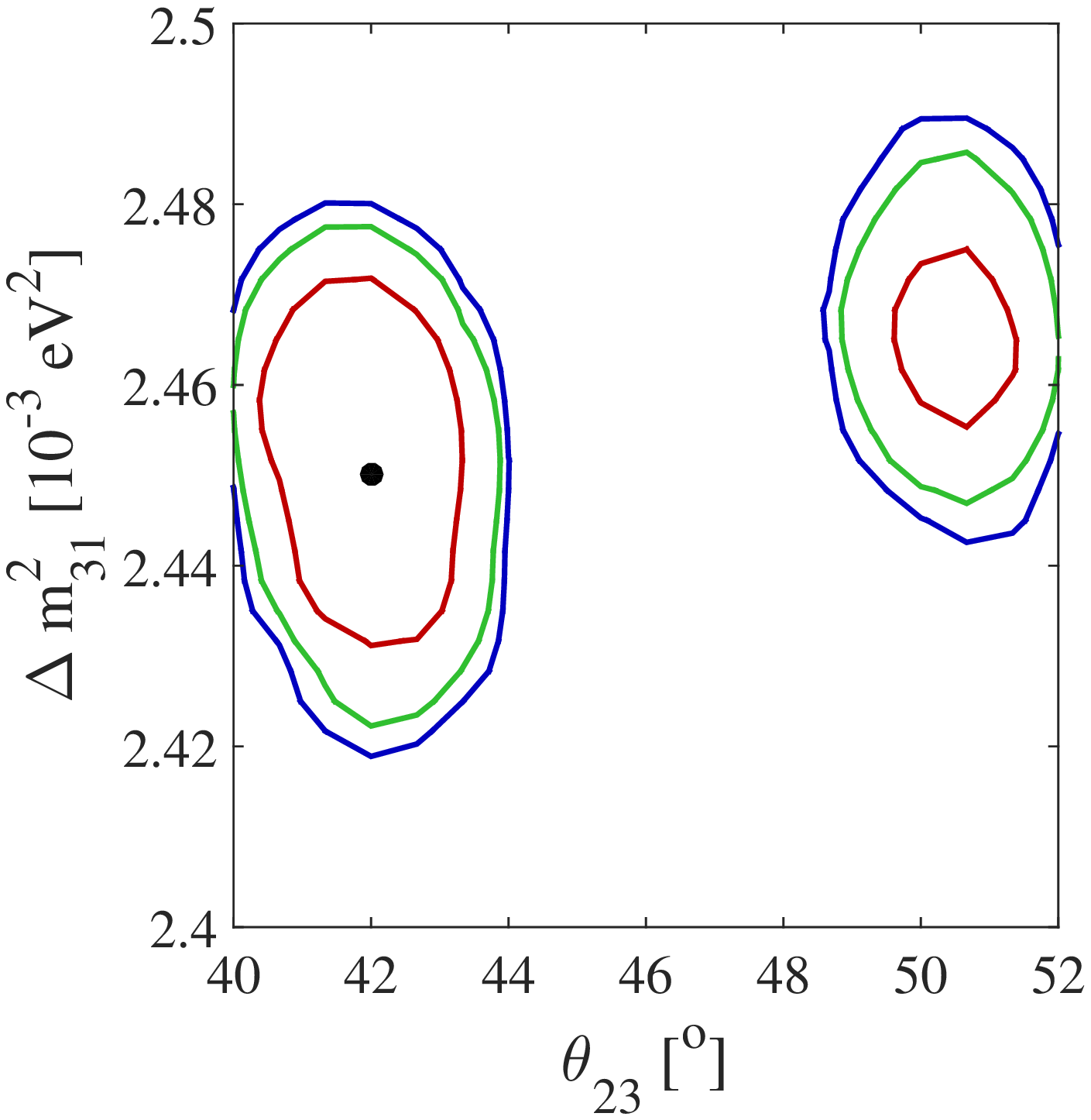, height=0.18\textheight, width=0.33\textwidth,
bbllx=0, bblly=0, bburx=480, bbury=440,clip=} &
\epsfig{file=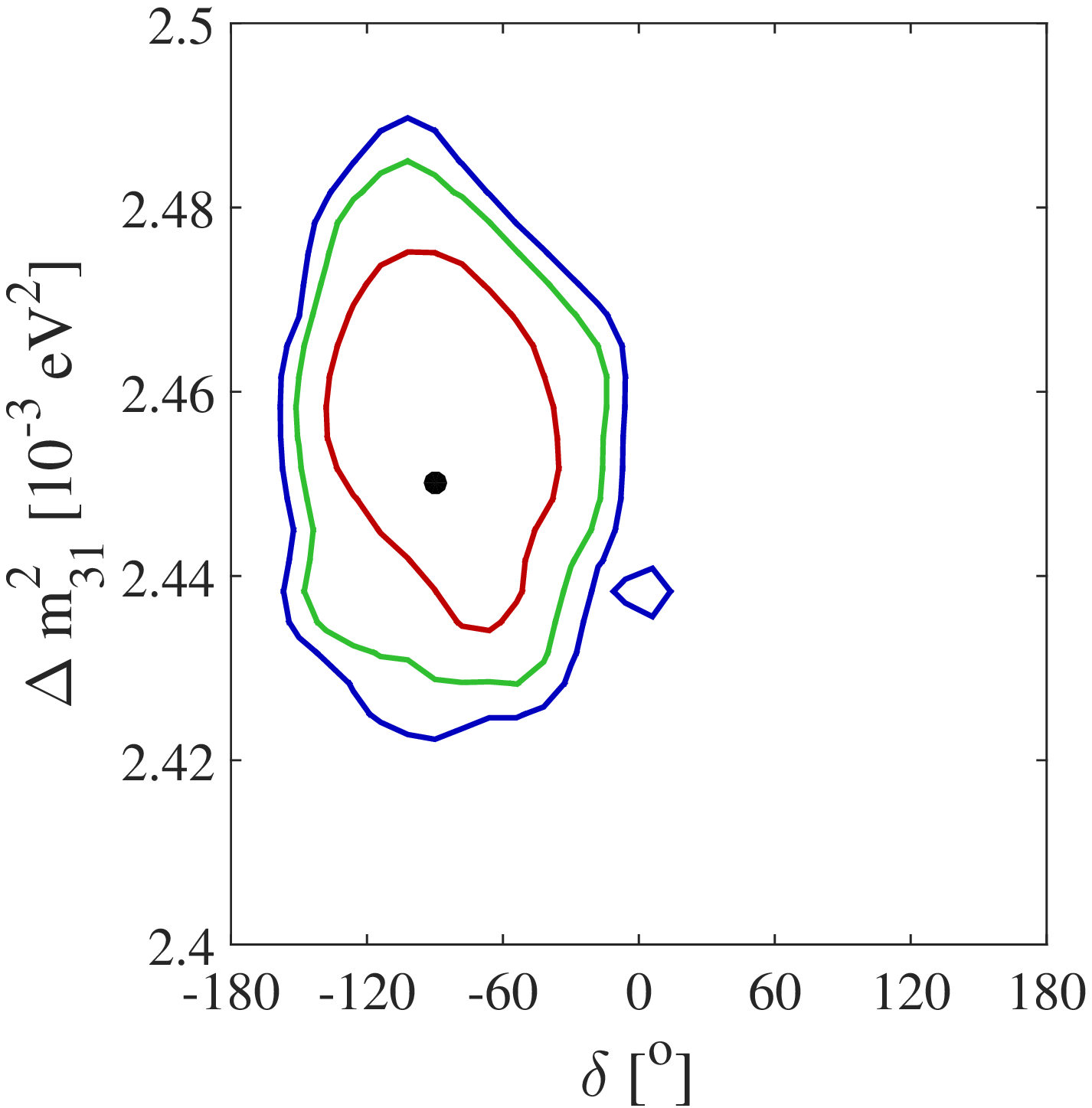, height=0.18\textheight, width=0.33\textwidth,
bbllx=0, bblly=0, bburx=480, bbury=440,clip=}
\\
\end{tabular}
\caption{\footnotesize Precision in the standard oscillation parameters in the 
presence of NSIs at DUNE. The contours shown correspond to 68~\% (red), 90~\%
(green) and 95~\% (blue) credible regions.}
\label{fig:standardtriangle}
\end{figure}

It becomes important to question whether the presence of NSIs will adversely
affect 
the precision measurement of these parameters or not. Many recent studies have
explored this 
question for
DUNE~\cite{Masud:2015xva,deGouvea:2015ndi,Coloma:2015kiu,Masud:2016bvp,
Masud:2016gcl} in the context of matter NSIs. In Fig.~\ref{fig:effect}, we show
the 
effect of NSIs on the precision measurement of $\theta_{23}$ and $\dcp$ when the
true 
values of these parameters are $42^\circ$ and $-90^\circ$, respectively. In the
left 
panel, we have set the true values of all NSI parameters to zero. We have then 
scanned the parameter space for four different cases, where the parameter space
consists 
of (a) only the standard oscillation parameters, (b) standard parameters and 
source/detector NSI parameters, (c) standard parameters and matter NSI
parameters 
and (d) standard parameters, source/detector NSI parameters and matter NSI
parameters. 
The results are displayed as different contours in the parameter space as the
90~\%~credible regions. This 
plot shows how the precision in $\theta_{23}$ and $\dcp$ changes as we change
our 
assumption about the parameter space when there are no NSIs in nature. 
We observe that the sensitivity in $\theta_{23}$ is not affected by scanning the
extra 
parameter space. Another significant feature is that the source/detector NSIs
do 
not play much of a role. This is expected, since the current bounds restrict
the 
allowed range of these parameters. There is some worsening of the sensitivity to
$\dcp$. 
In the right panel, we have given a similar plot, but with non-zero values of
the NSI 
parameters. This plot shows the successive worsening of precision as more NSIs 
are introduced. 
The innermost contour displays the allowed region, where NSIs are present in
nature, 
but we only choose to marginalize over the standard parameters. The sensitivity
to 
the standard parameters obtained from such an analysis would be erroneously
optimistic.  
In order to see the actual worsening of sensitivity because of the presence of
NSIs, 
we compare the dotted black curve of the left panel (zero NSIs, standard
oscillation scenario) 
with the solid red curve of the left panel (non-zero NSIs, all NSI parameters
included in 
the fit). We find that the error in the measurement of $\dcp$ (the size in
$\delta$ of the 90~\% credible region) increases 
from around $90^\circ$ to almost $180^\circ$. In addition, we find a degenerate
solution 
in $\theta_{23}$ in the wrong octant. These additional degeneracies have been
studied 
recently in Refs.~\cite{deGouvea:2015ndi,Coloma:2015kiu}.

For completeness, we compute the precision in $\theta_{13}$, $\theta_{23}$,
$\dcp$ and 
$\dm{31}$ that DUNE will reach in the presence of NSIs. The results are shown
in 
Fig.~\ref{fig:standardtriangle}, where a striking feature is the 
appearance of a degenerate solution in $\theta_{23}$.

\subsection{Constraining NSI parameters at DUNE}

So far, we have discussed the effect of NSIs on the measurement of the standard 
oscillation parameters. In addition, DUNE can place bounds on the NSI 
parameters due to its high statistics. In order to estimate these projected 
bounds from DUNE, we set the true values of the NSI parameters to zero.
Varying the fit values, we construct the 90~\% credible region for the 
value of the parameter placed by DUNE. 
These 90~\% credible regions are obtained by marginalizing over all the 
other parameters. In Table~\ref{tab:bounds}, we list the 90~\% credible upper
bounds that DUNE 
can impose. We have performed the computations for three different cases: 
(a) the only NSIs are source/detector NSIs, (b) the only NSIs are 
matter NSIs and (c) all NSIs exist simultaneously. 

\begin{table}[htb]
\begin{center}
 \begin{tabular}{|c|c|c|c|c|}
  \hline\hline
  Parameter & Only source/detector NSIs & Only matter NSIs & All NSIs & Current
bound \\
  \hline\hline
  $|\epssme|$ & 0.017 & & 0.022 & 0.026 \\
  $|\epssmm|$ & 0.070 & & 0.065 & 0.078 \\
  $|\epssmt|$ & 0.009 & & 0.014 & 0.013 \\
  \hline
  $|\epsdme|$ & 0.021 & & 0.023 & 0.025 \\
  $|\epsdte|$ & 0.028 & & 0.035 & 0.041 \\
  \hline
  ${\epsmee}'$ & & $(-0.7,+0.8)$ & $(-0.8,+0.9)$ & $(-4.2,+4.2)$ \\
  $|\epsmme|$ & & 0.051 & 0.074 & 0.330 \\
  $|\epsmte|$ & & 0.17 & 0.19 & 3.00 \\
  $|\epsmtm|$ & & 0.031 & 0.038 & 0.040 \\
  ${\epsmtt}'$ & & $(-0.08,+0.08)$ & $(-0.08,+0.08)$ & $(-0.15,+0.15)$ \\
  \hline
   \end{tabular}
\end{center}
 \caption{\footnotesize Expected 90~\% credible regions on NSI parameters from
DUNE.}
 \label{tab:bounds}
\end{table}

In general, investigating Table~\ref{tab:bounds}, we see that in going from the
case of (a) only source/detector NSIs or (b) only matter
NSIs to the case of (c) both source/detector and matter NSIs, the bounds imposed
by DUNE get
weaker. This is expected because of the expansion of the parameter
space. 
(For $|\epssmm|$, the precision appears to improve marginally after the inclusion of 
all NSIs. This is merely an artifact of the randomness inherent in the 
Monte Carlo simulation and should be taken with a pinch of salt. The relative difference 
between the two numbers is small enough for them to be practically 
equal within the precision of our Monte Carlo simulation.)
Using all
NSIs, we find that all bounds are improved or basically the same as the current
bounds. We also
see that the most general bounds imposed on the source/detector NSI parameters
are only slightly
better than the existing bounds. This shows that the main contribution to the
sensitivity to these
parameters comes from the prior introduced for them. Data from DUNE itself
contribute only 
to the extent of providing more statistics without any significant physics
advantage.  On the other hand, we find that the bounds on matter NSI parameters
are
made substantially more stringent than the existing bounds. In particular, the
bounds on ${\epsmee}'$, $|\epsmme|$ and $|\epsmte|$ are
improved by a factor of around five to 15, whereas the bounds on $|\epsmtm|$ and
${\epsmtt}'$ are
more or less the same. Our results on the bounds on the matter NSI parameters
are consistent with the ones obtained in Ref.~\cite{Coloma:2015kiu}. 
 It is worth pointing out that the current bounds on the NSI parameters were
derived assuming the existence of only one NSI parameter at a time, whereas we
have obtained our bounds by allowing all relevant parameters to vary at the same
time. 

\subsection{Correlations between source/detector and matter NSIs}

\begin{figure}
\begin{tabular}{cc}
\epsfig{file=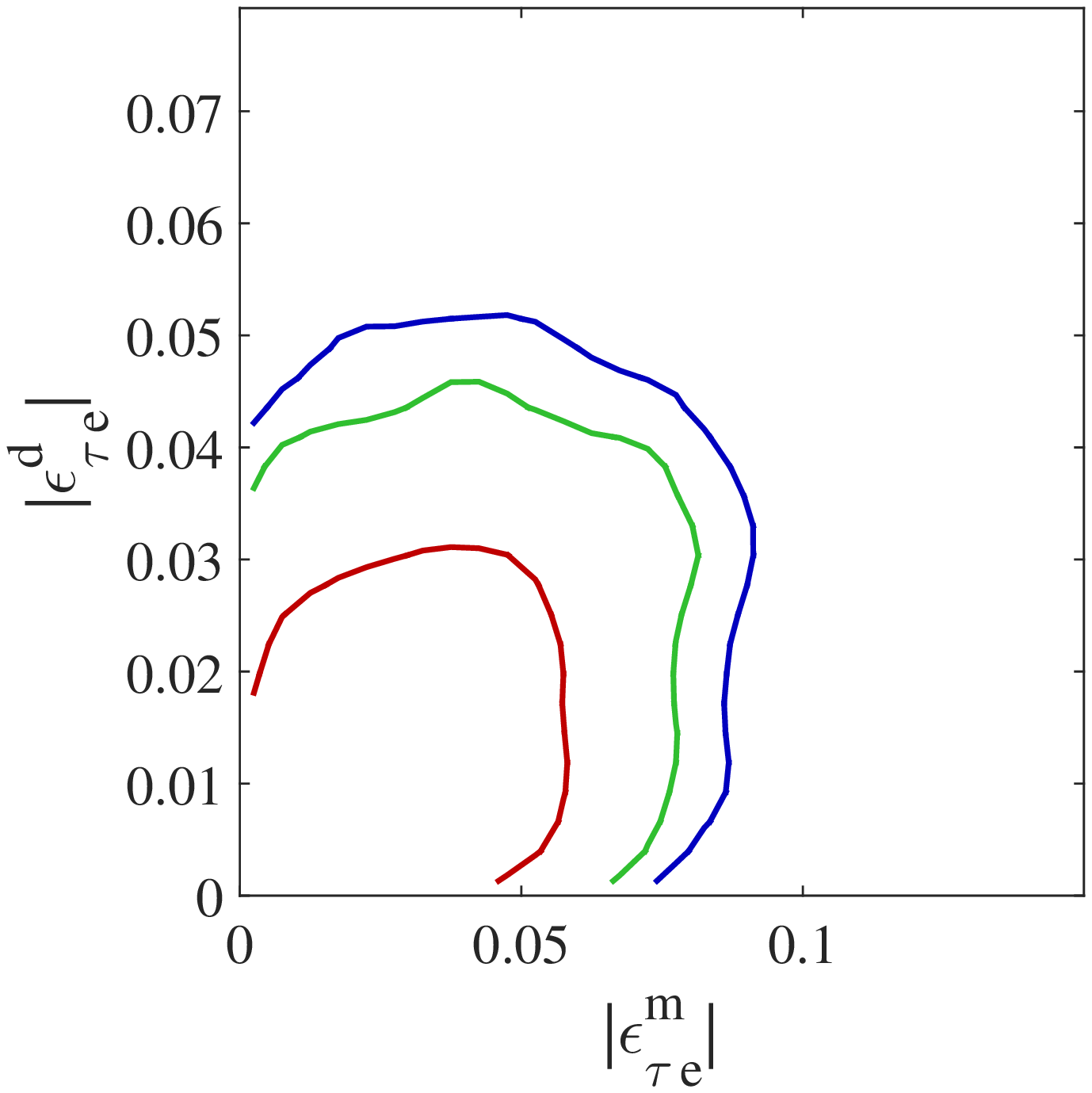, width=0.5\textwidth, bbllx=0, bblly=0,
bburx=420, bbury=420,clip=} &
\epsfig{file=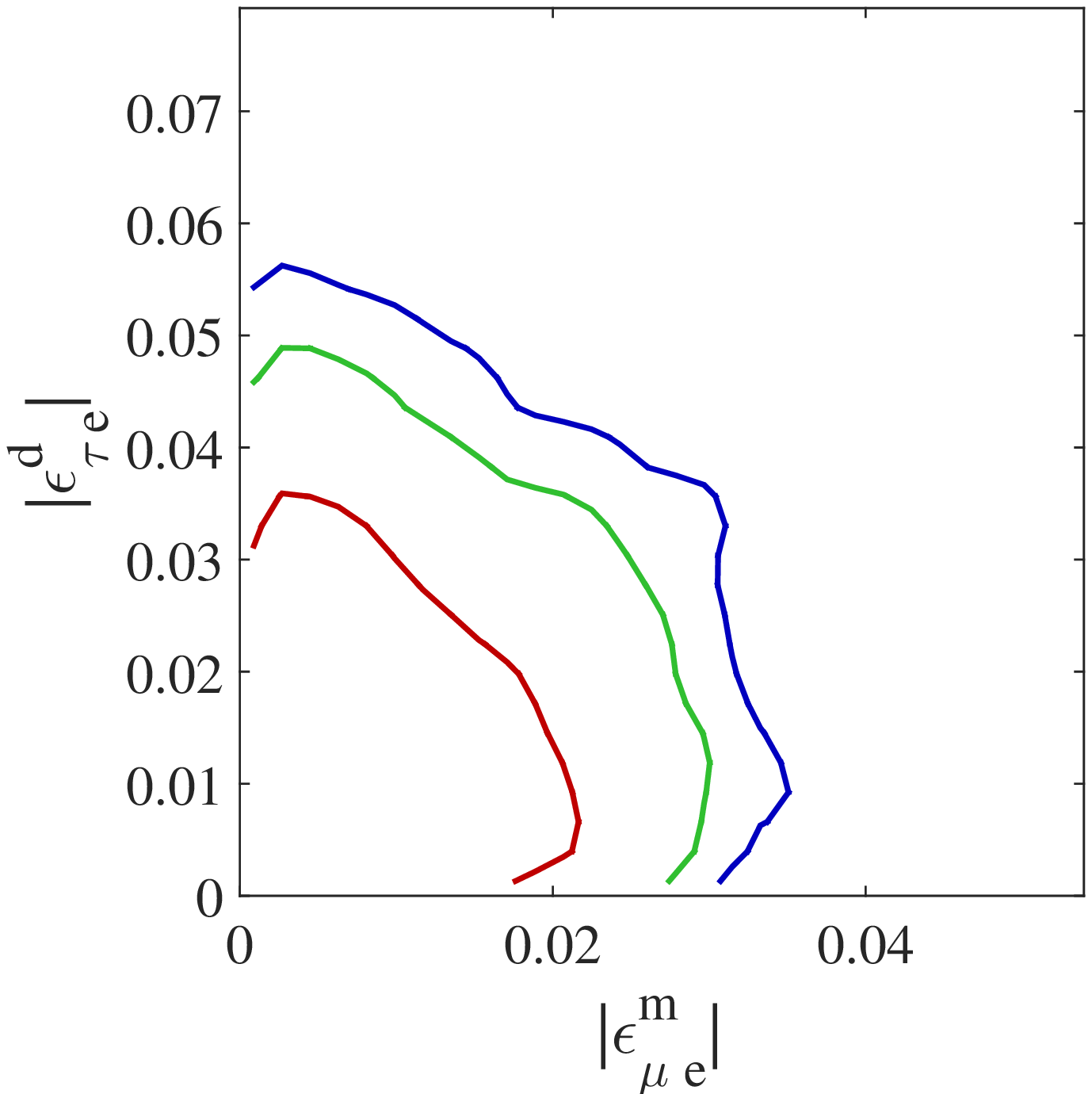, width=0.5\textwidth, bbllx=0, bblly=0,
bburx=420, bbury=420,clip=} 
\end{tabular}
\caption{\footnotesize Correlations between matter NSI parameters and
source/detector 
NSI parameters at DUNE. The 68~\% (red), 90~\% (green) and 95~\% (blue) credible
regions are shown in the 
$\epsmte-\epsdte$ plane in the left panel and in the $\epsmme-\epsdte$ 
plane in the right panel.}
\label{fig:correl}
\end{figure}

\begin{figure}
\begin{tabular}{ccc}
\epsfig{file=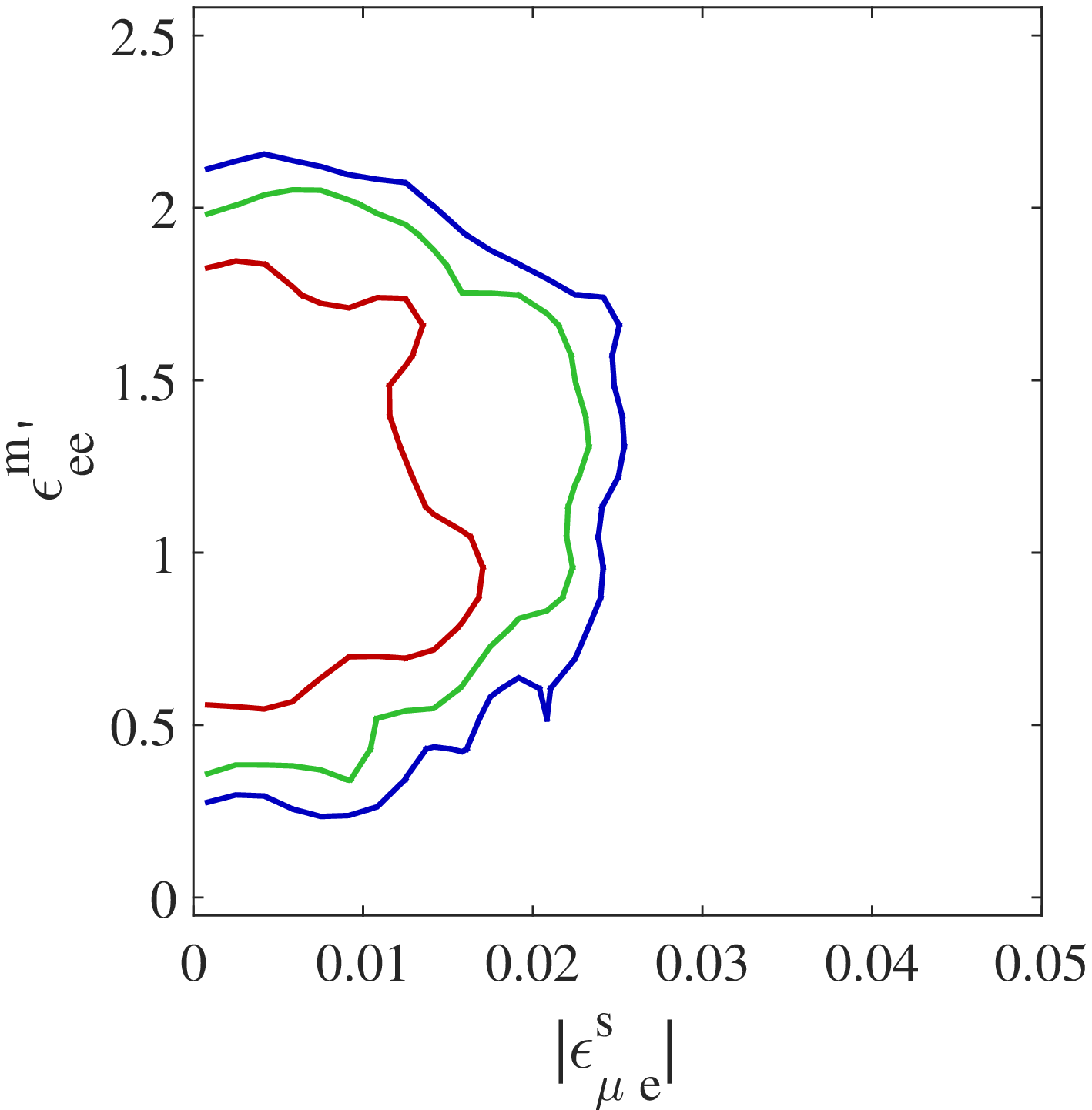, height=0.18\textheight, width=0.33\textwidth, bbllx=0,
bblly=0, bburx=480, bbury=440,clip=} &
\epsfig{file=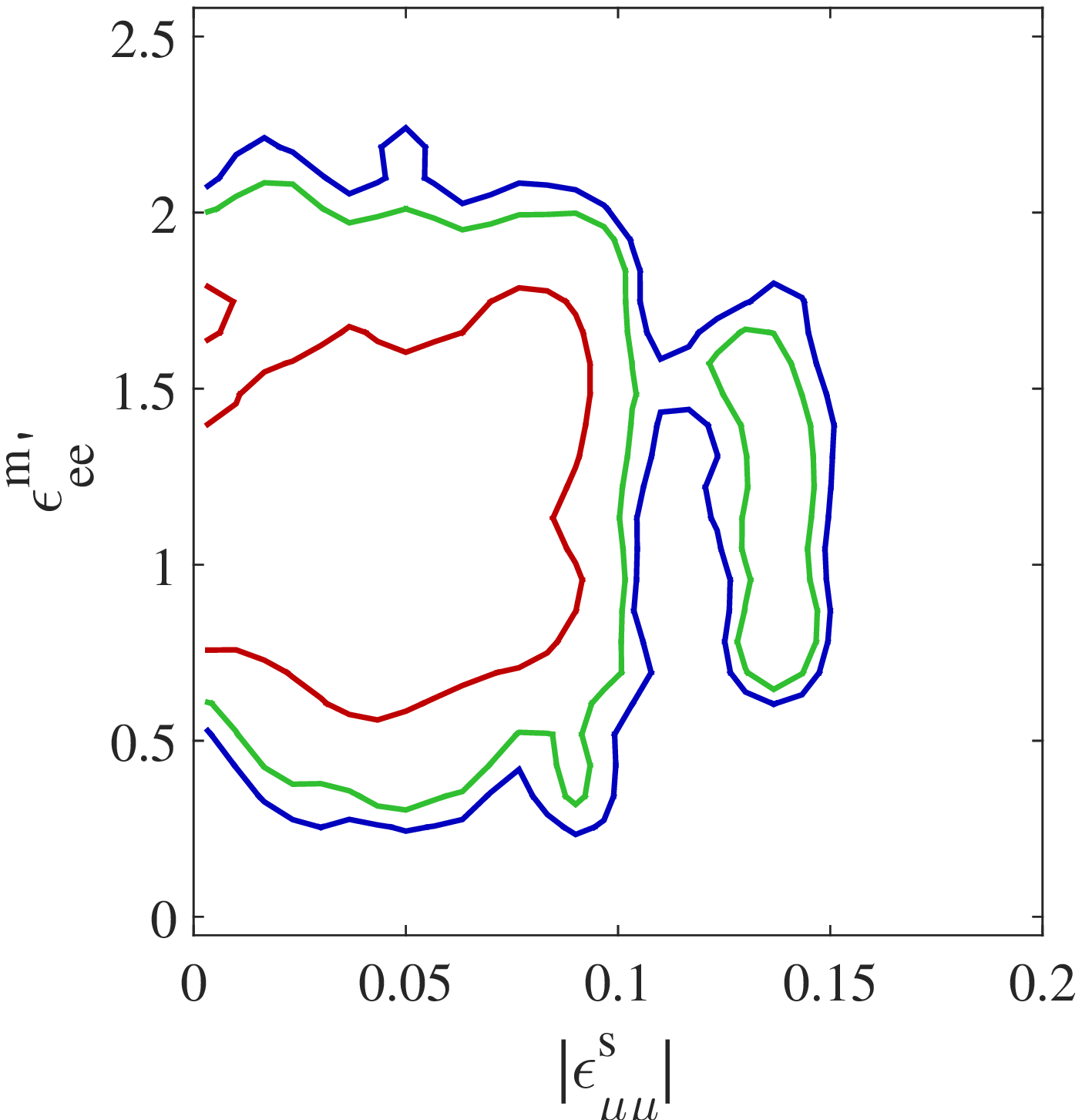, height=0.18\textheight, width=0.33\textwidth, bbllx=0,
bblly=0, bburx=480, bbury=440,clip=} & 
\epsfig{file=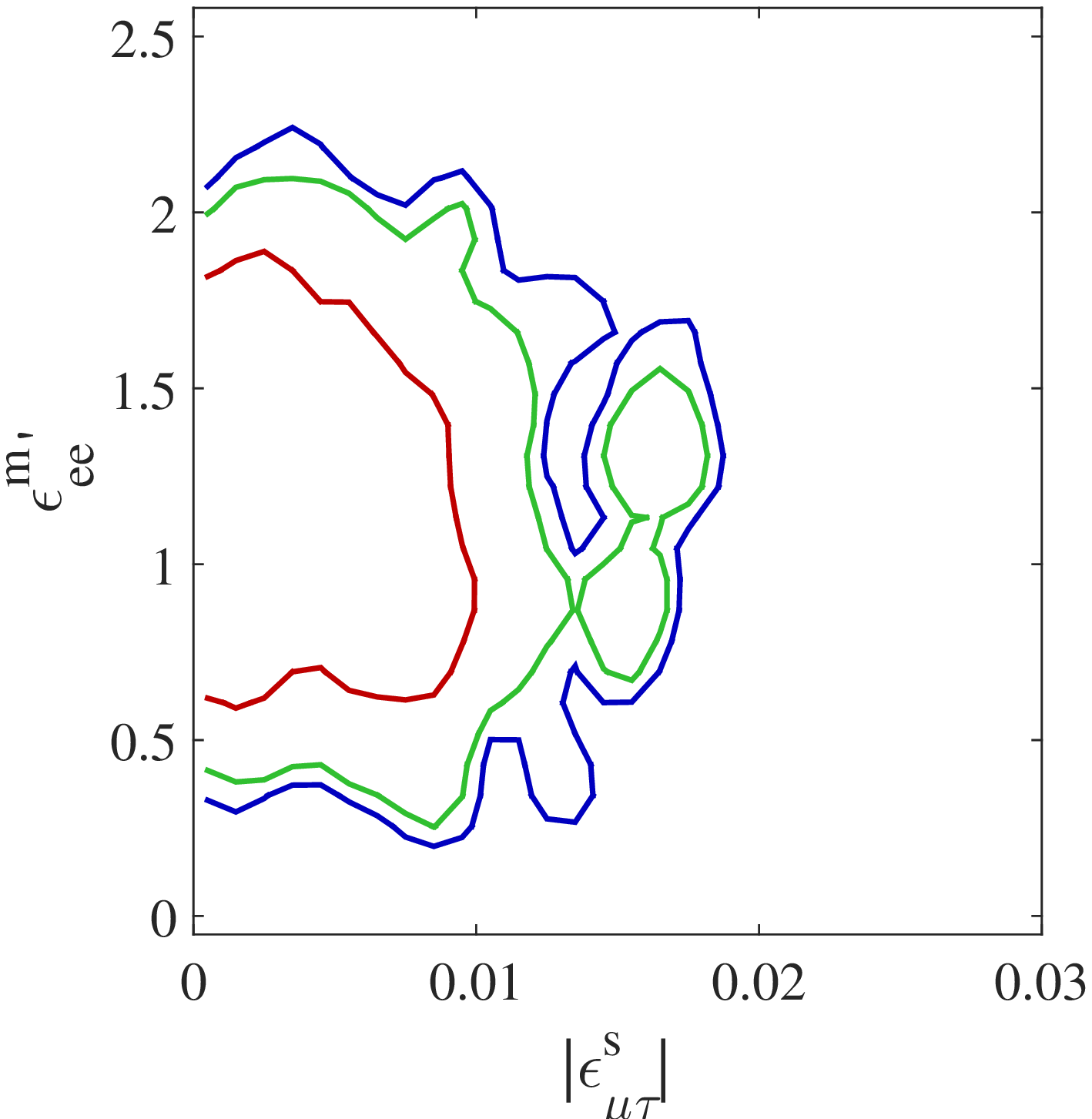, height=0.18\textheight, width=0.33\textwidth, bbllx=0,
bblly=0, bburx=480, bbury=440,clip=} \\ 
\epsfig{file=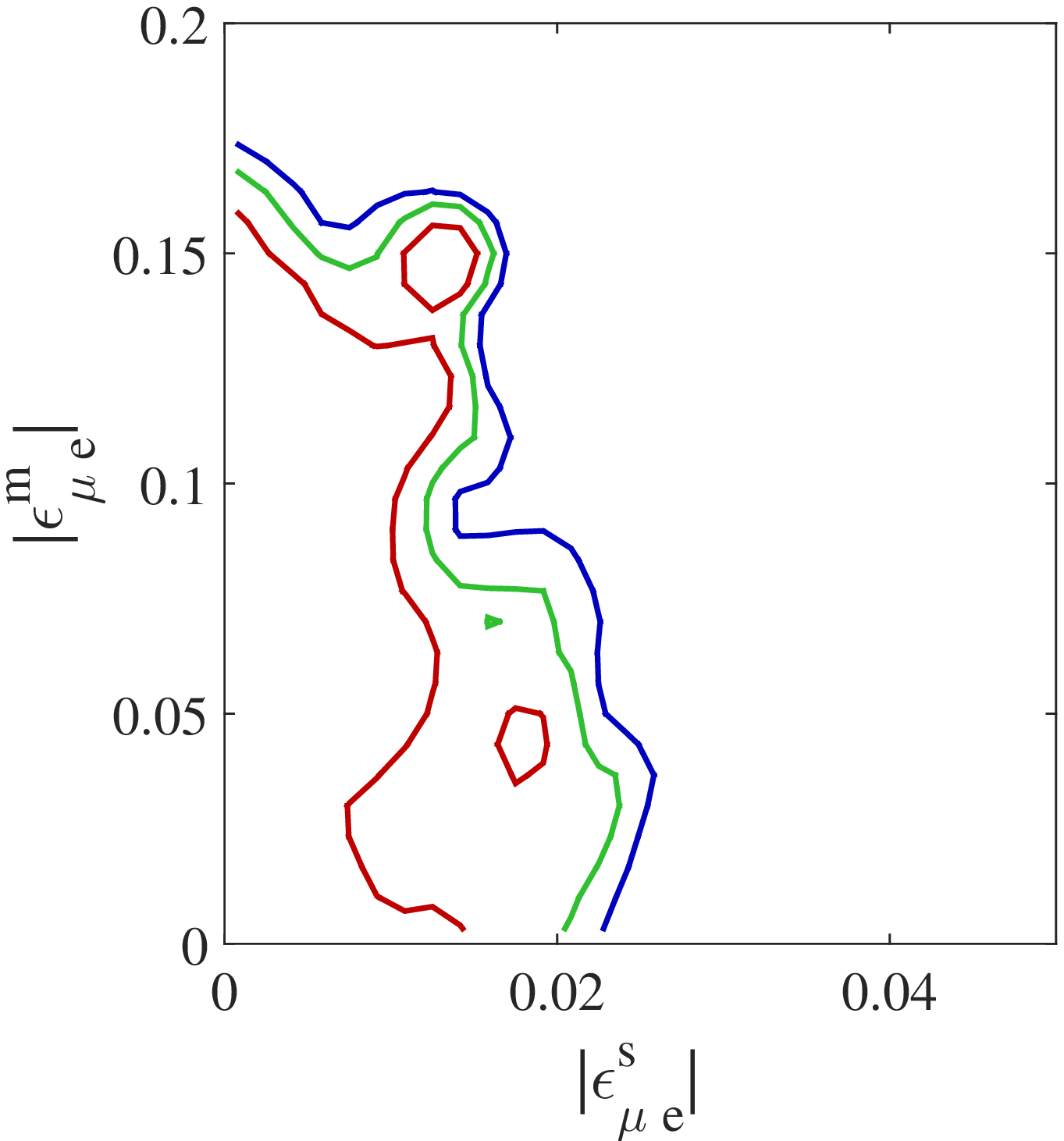, height=0.18\textheight, width=0.33\textwidth, bbllx=0,
bblly=0, bburx=480, bbury=440,clip=} &
\epsfig{file=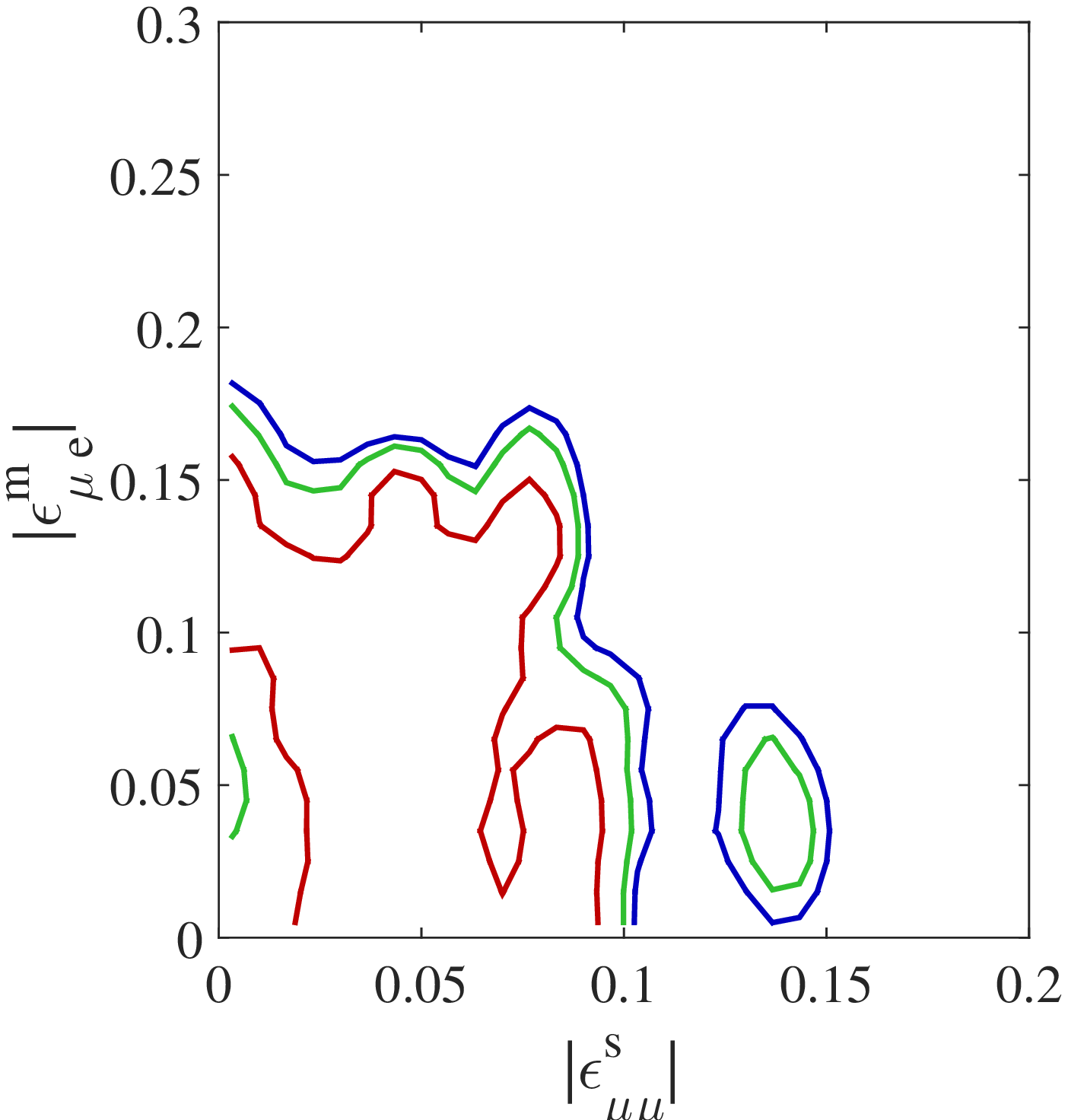, height=0.18\textheight, width=0.33\textwidth, bbllx=0,
bblly=0, bburx=480, bbury=440,clip=} & 
\epsfig{file=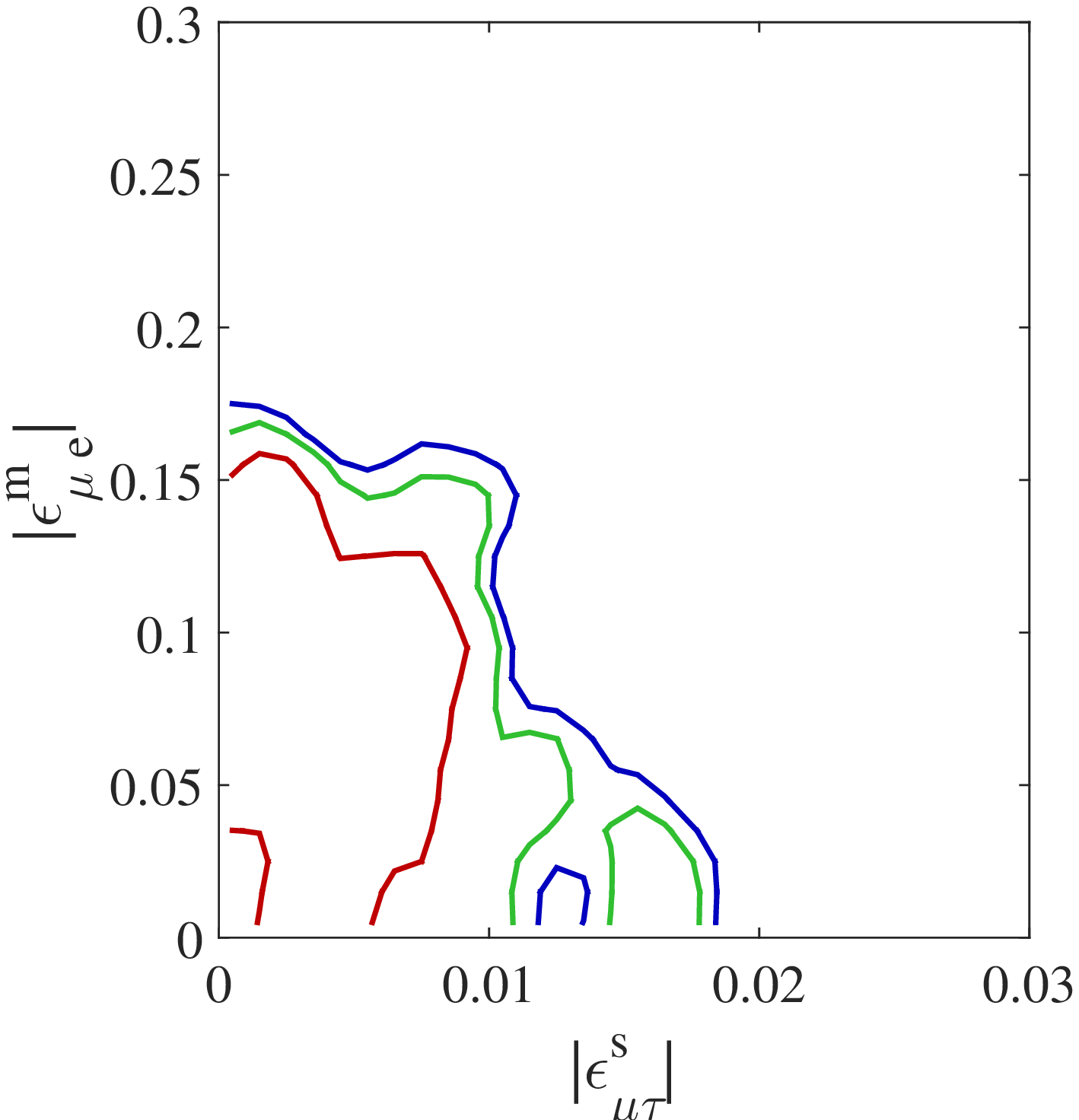, height=0.18\textheight, width=0.33\textwidth, bbllx=0,
bblly=0, bburx=480, bbury=440,clip=} \\ 
\epsfig{file=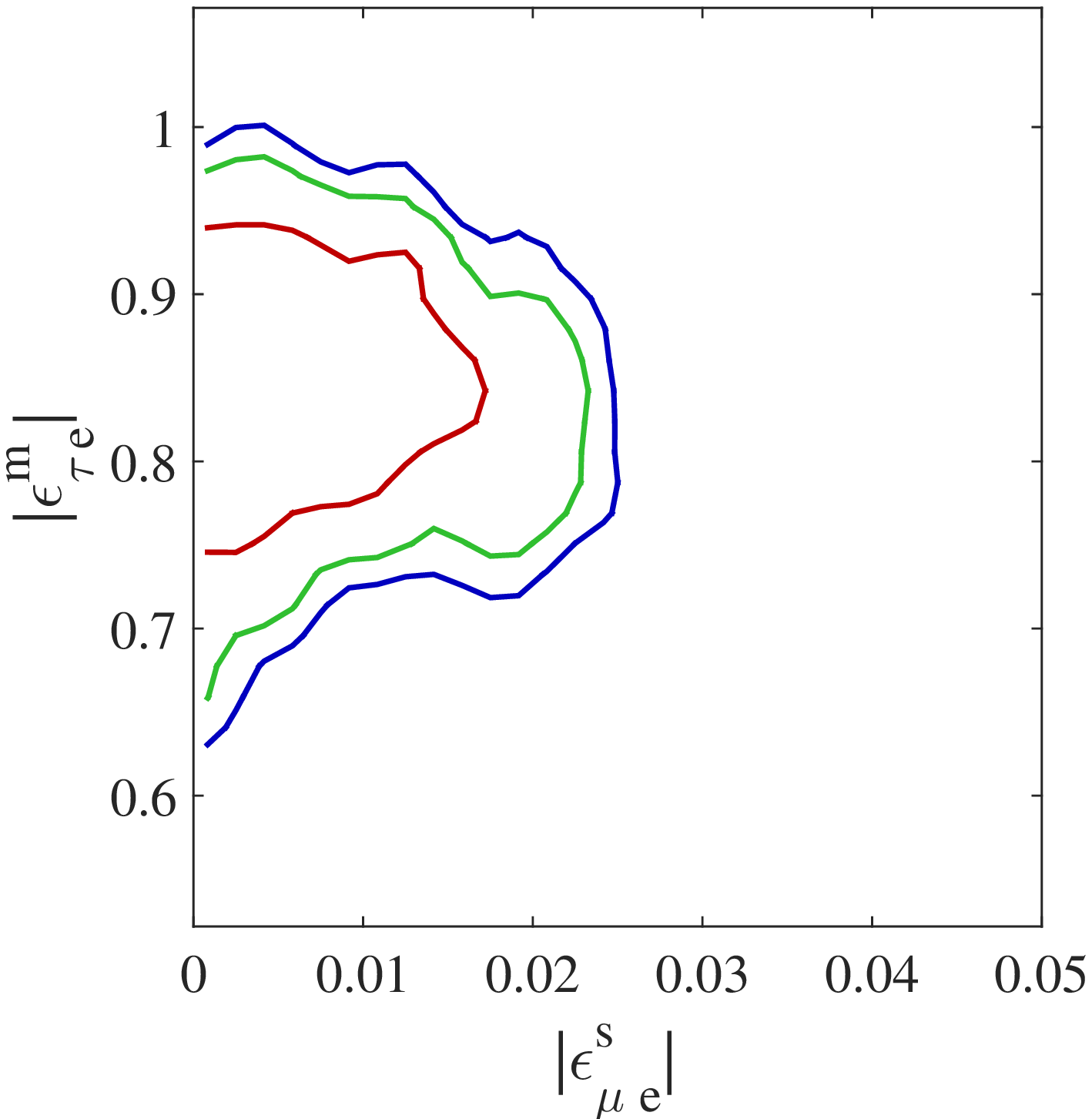, height=0.18\textheight, width=0.33\textwidth, bbllx=0,
bblly=0, bburx=480, bbury=440,clip=} &
\epsfig{file=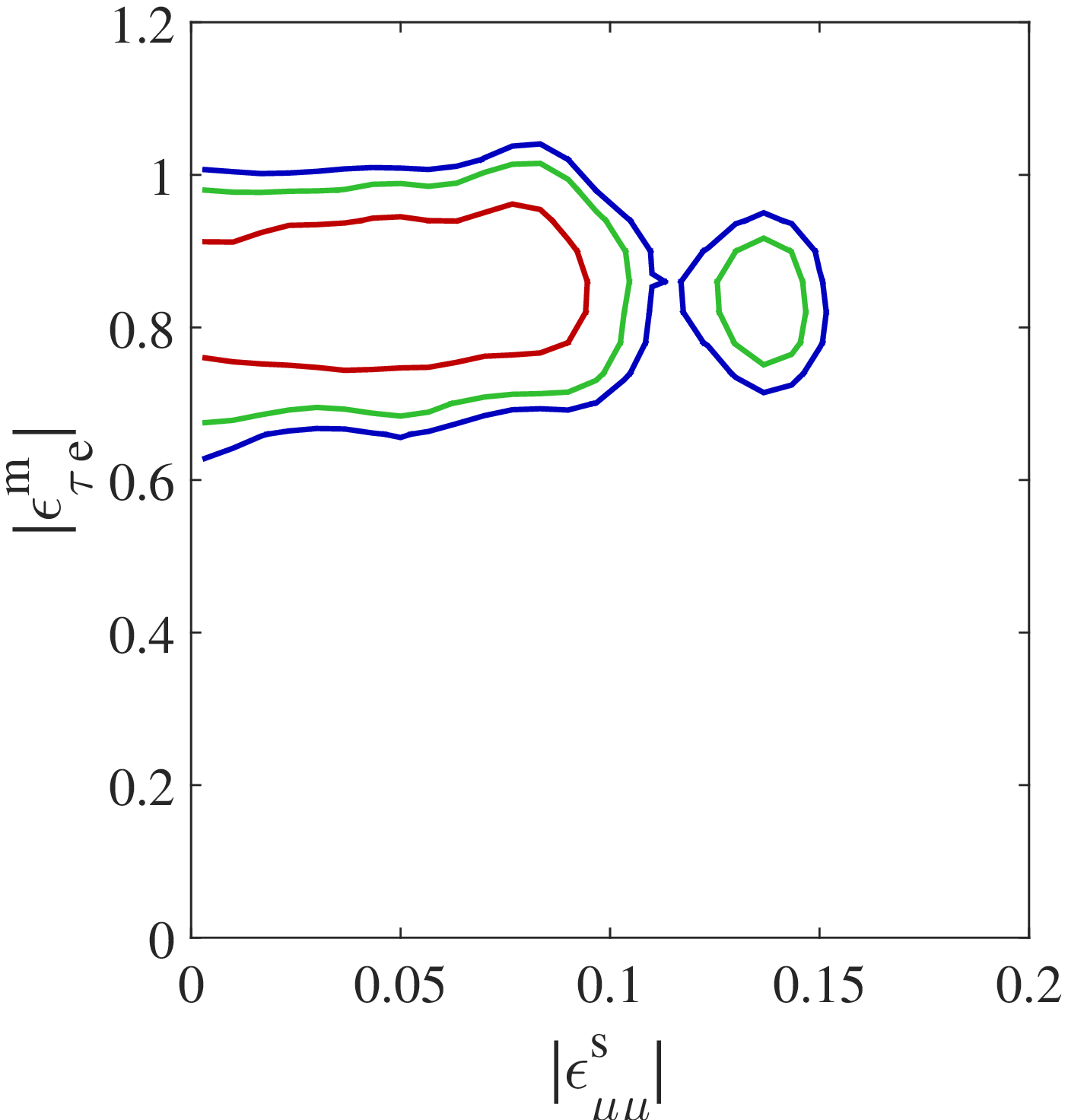, height=0.18\textheight, width=0.33\textwidth, bbllx=0,
bblly=0, bburx=480, bbury=440,clip=} & 
\epsfig{file=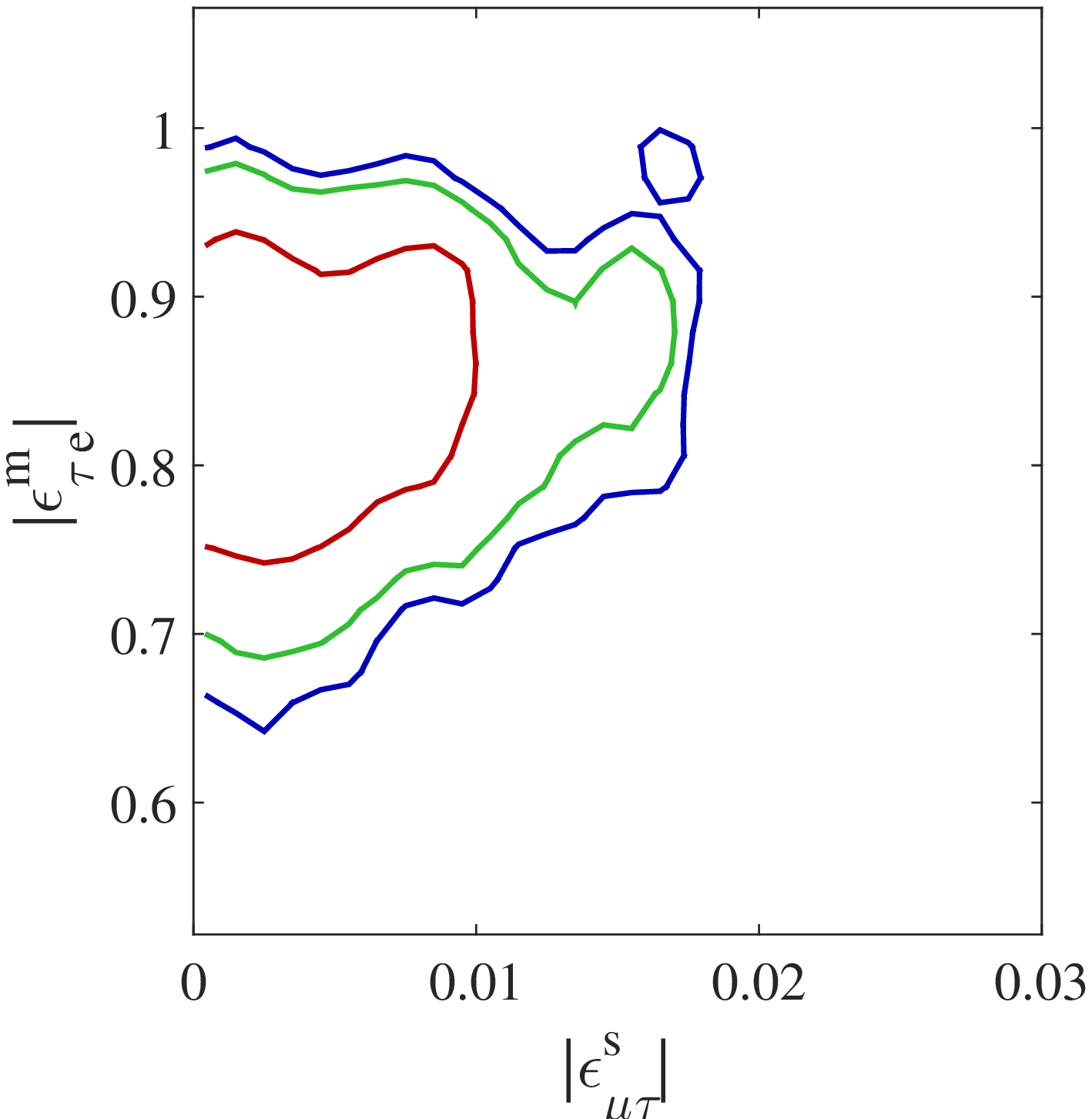, height=0.18\textheight, width=0.33\textwidth, bbllx=0,
bblly=0, bburx=480, bbury=440,clip=} \\ 
\epsfig{file=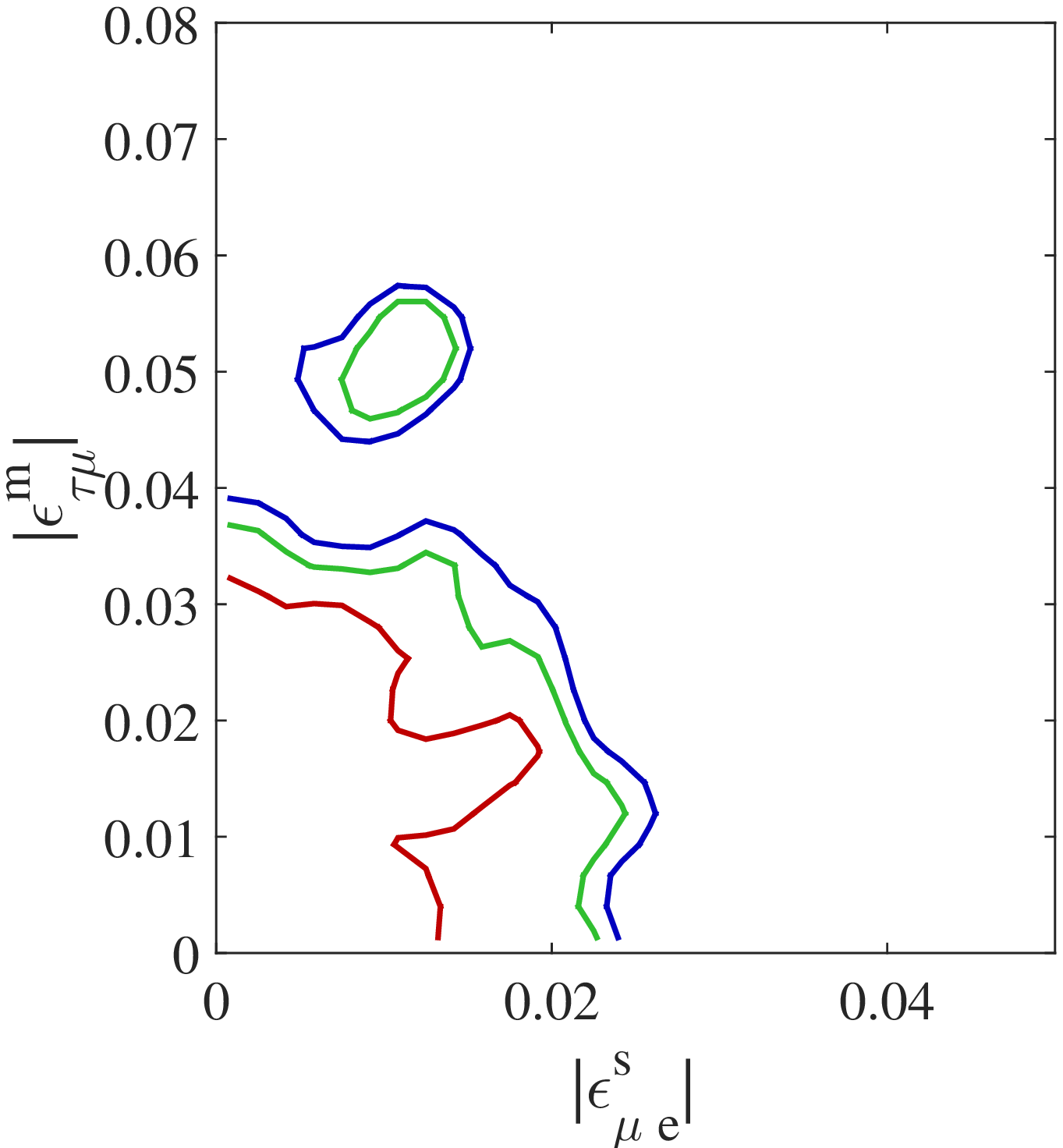, height=0.18\textheight, width=0.33\textwidth, bbllx=0,
bblly=0, bburx=480, bbury=440,clip=} &
\epsfig{file=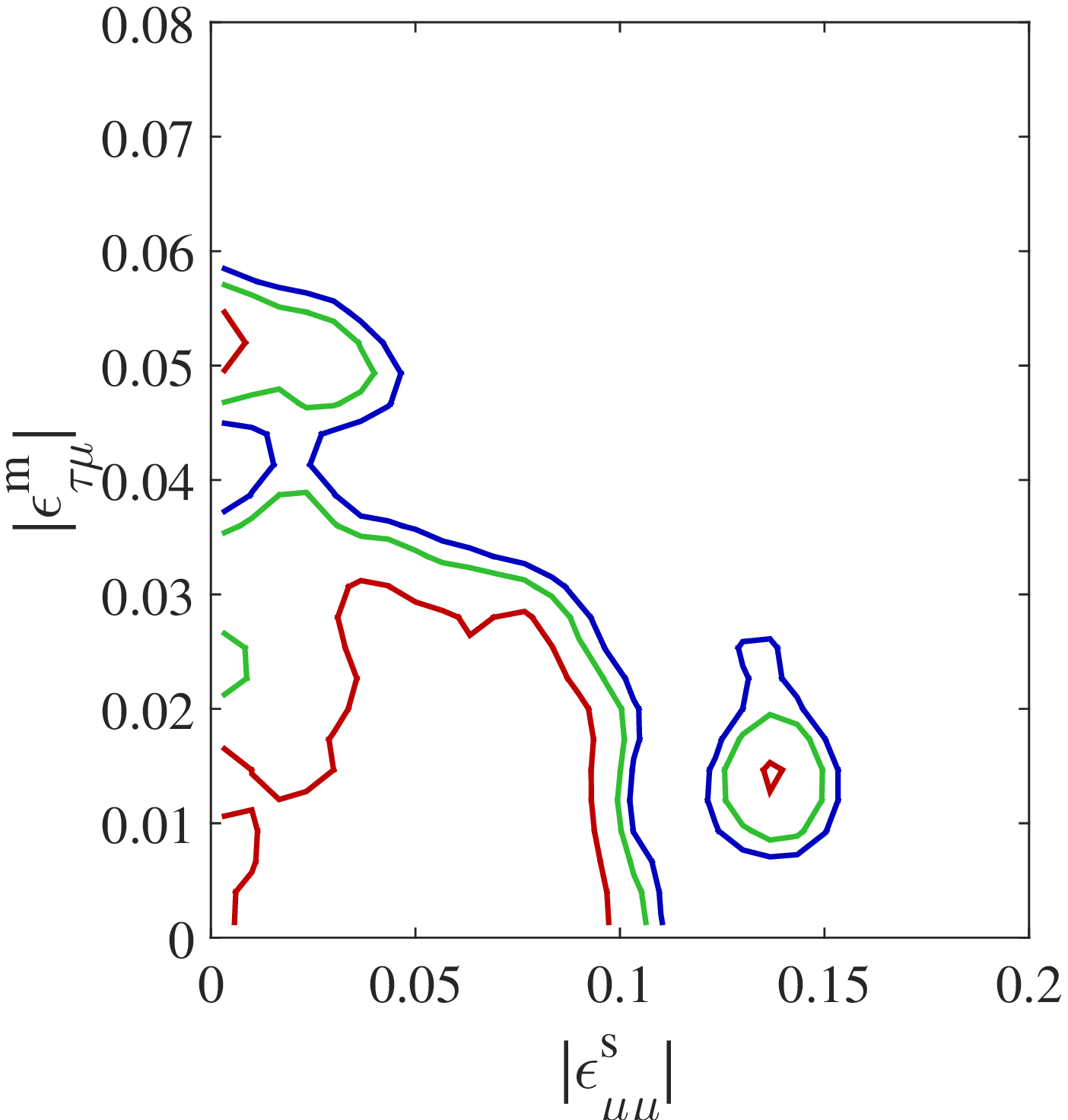, height=0.18\textheight, width=0.33\textwidth, bbllx=0,
bblly=0, bburx=480, bbury=440,clip=} & 
\epsfig{file=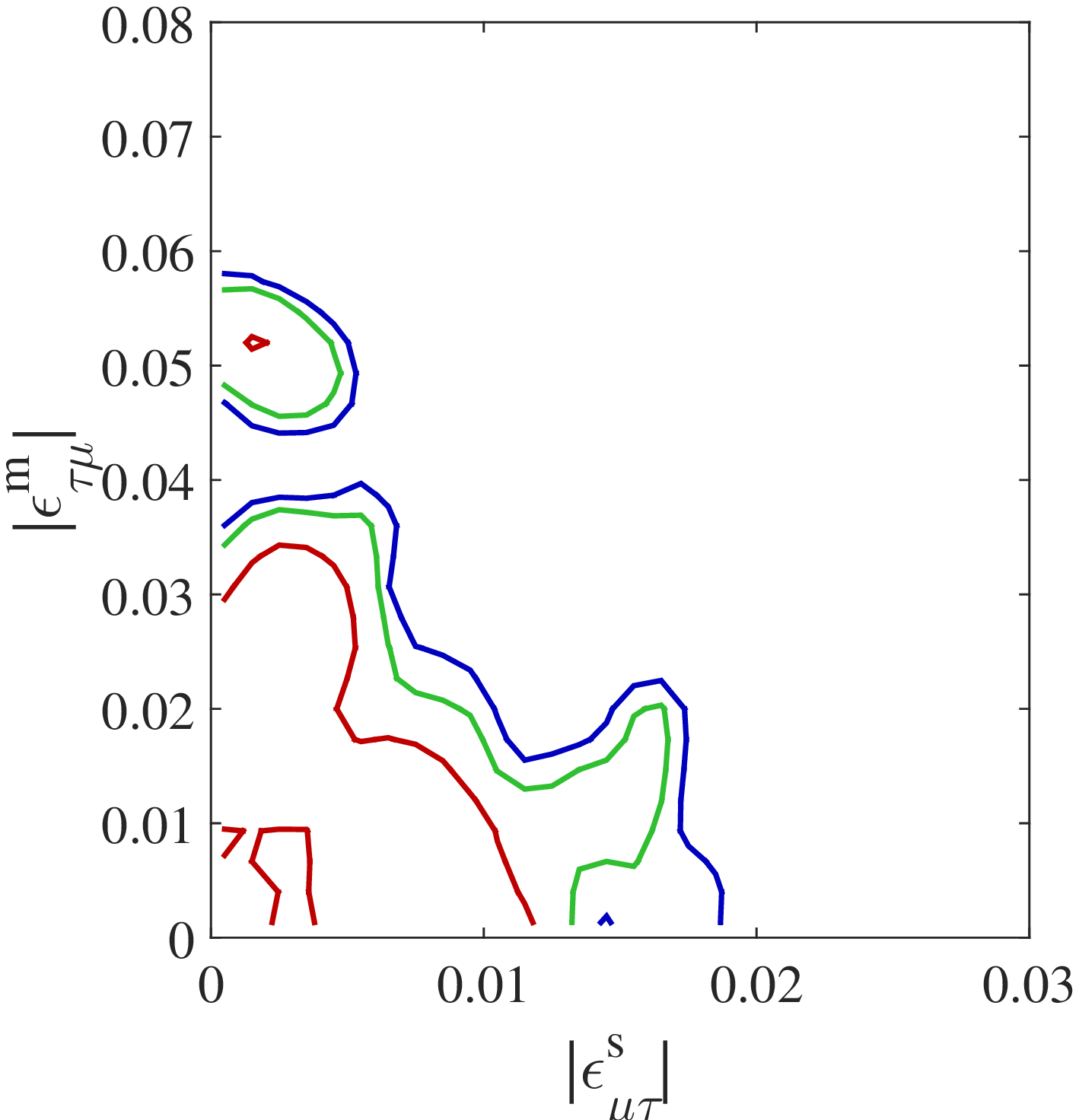, height=0.18\textheight, width=0.33\textwidth, bbllx=0,
bblly=0, bburx=480, bbury=440,clip=} \\ 
\epsfig{file=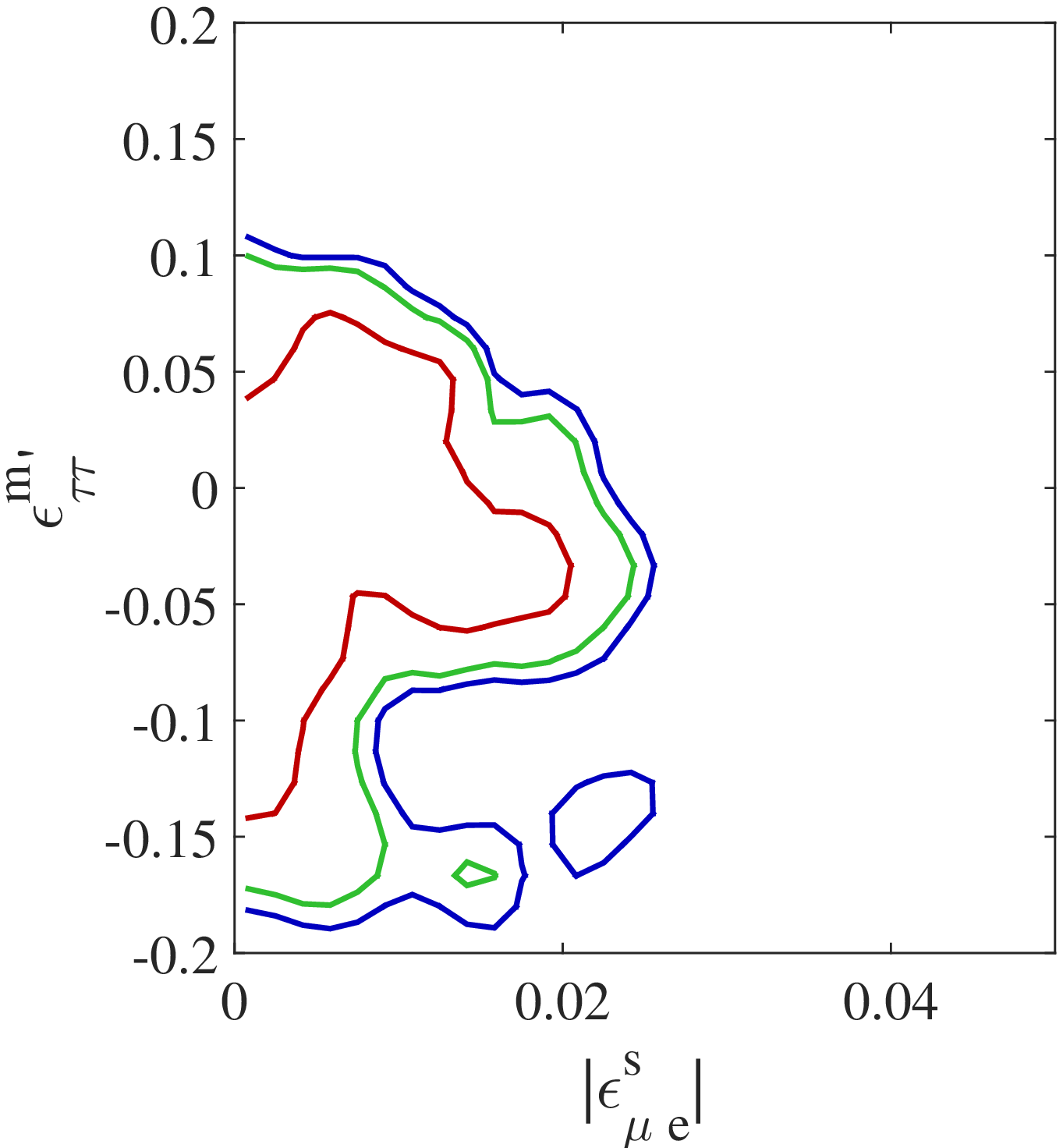, height=0.18\textheight, width=0.33\textwidth, bbllx=0,
bblly=0, bburx=480, bbury=440,clip=} &
\epsfig{file=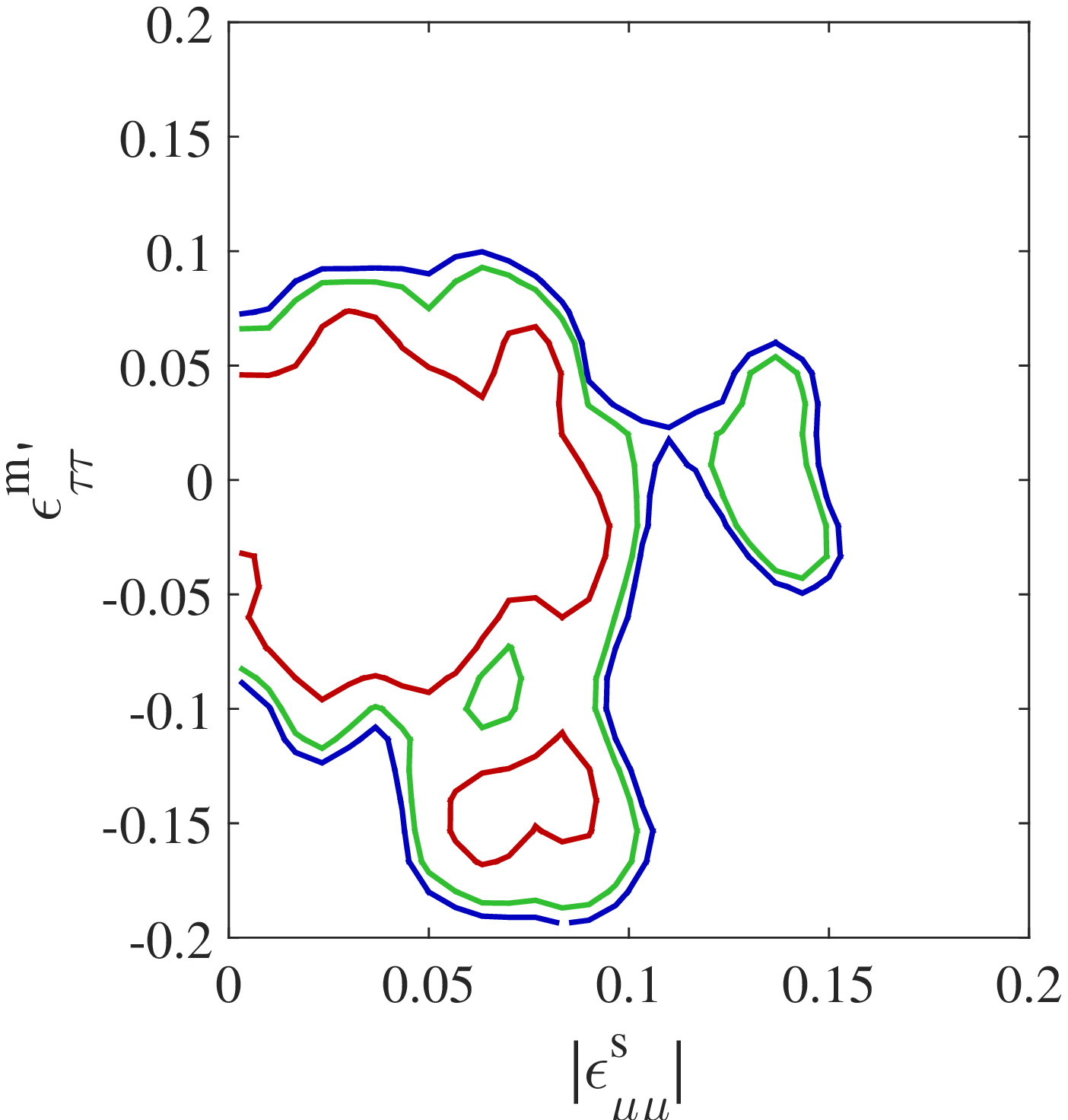, height=0.18\textheight, width=0.33\textwidth, bbllx=0,
bblly=0, bburx=480, bbury=440,clip=} & 
\epsfig{file=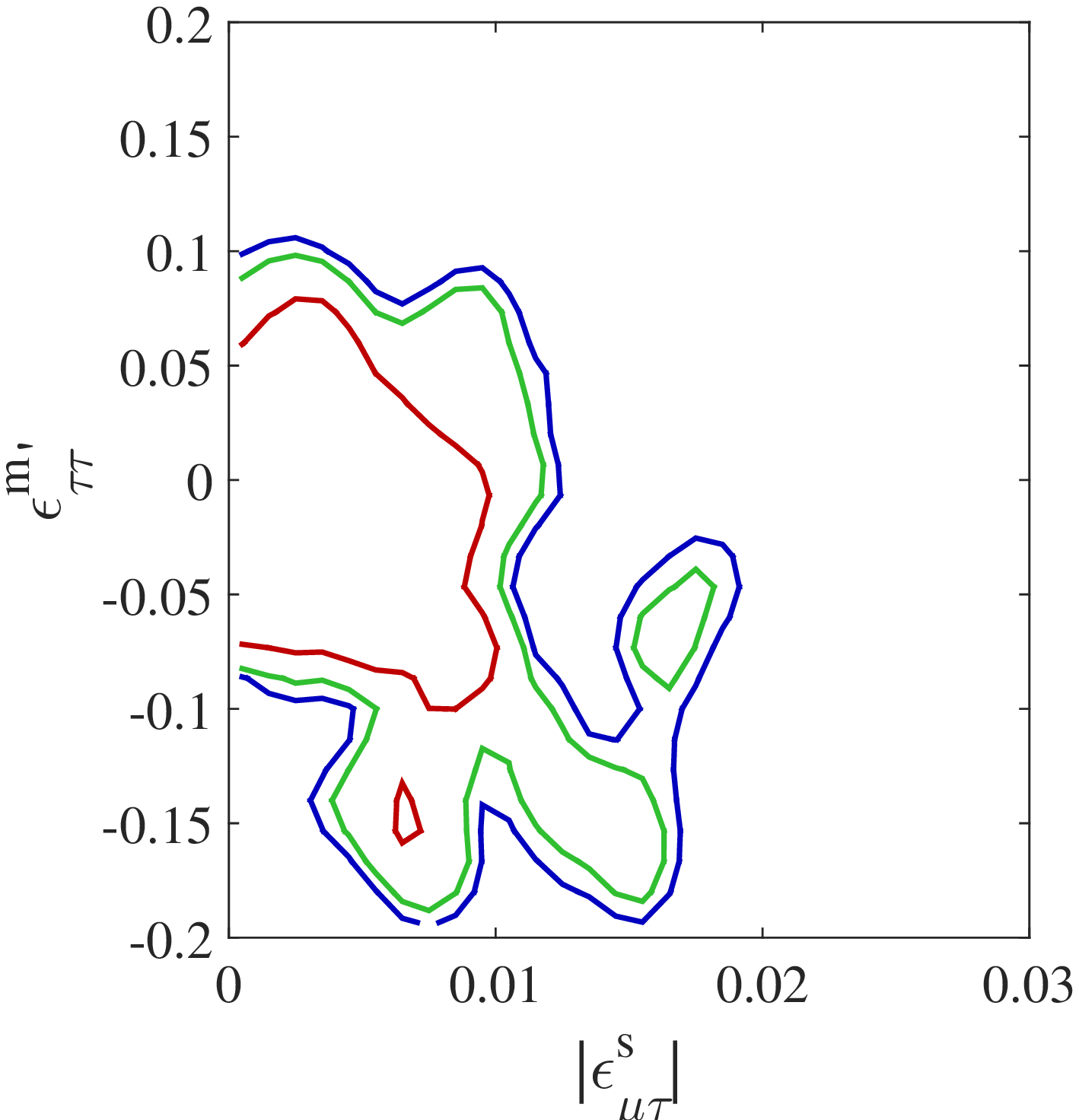, height=0.18\textheight, width=0.33\textwidth, bbllx=0,
bblly=0, bburx=480, bbury=440,clip=} 
\end{tabular}
\caption{\footnotesize Correlations between matter NSI parameters and source 
NSI parameters with current bounds at DUNE. The 68~\% (red), 90~\% (green) and
95~\% (blue) credible regions are shown.}
\label{fig:allcorrelss}
\end{figure}

\begin{figure}
\begin{tabular}{cc}
\epsfig{file=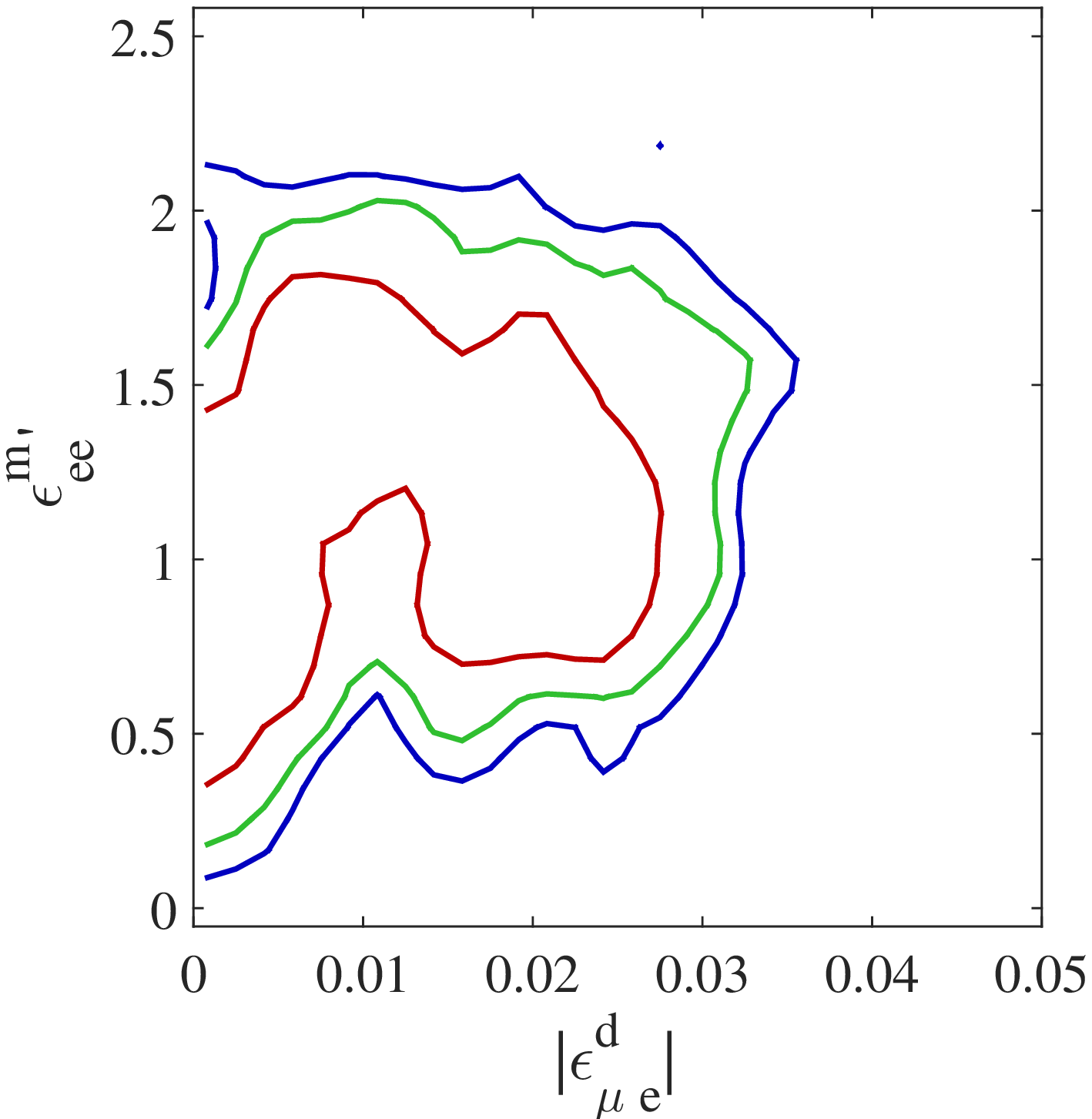, height=0.18\textheight, width=0.33\textwidth, bbllx=0,
bblly=0, bburx=480, bbury=440,clip=} &
\epsfig{file=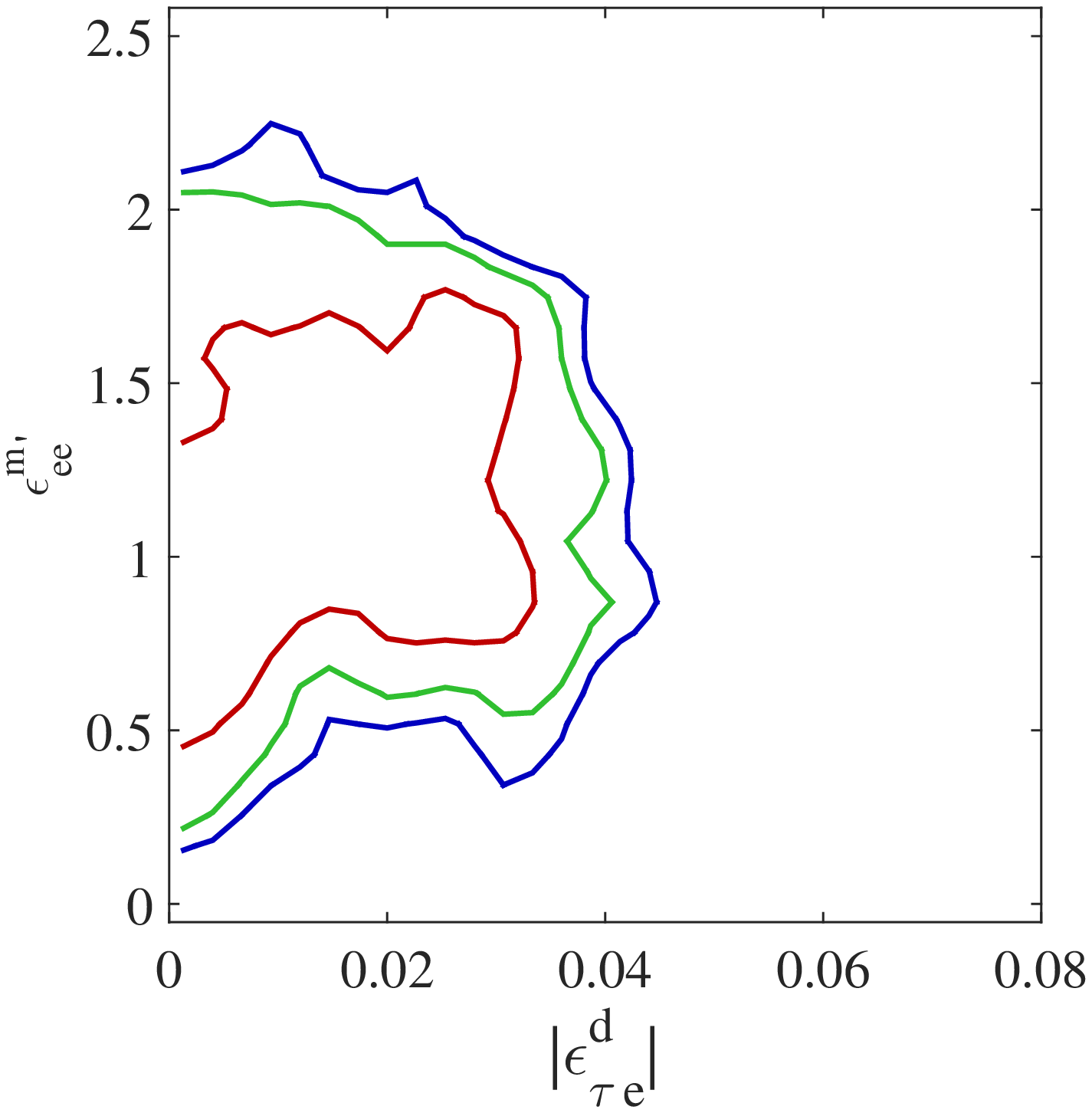, height=0.18\textheight, width=0.33\textwidth, bbllx=0,
bblly=0, bburx=480, bbury=440,clip=} \\ 
\epsfig{file=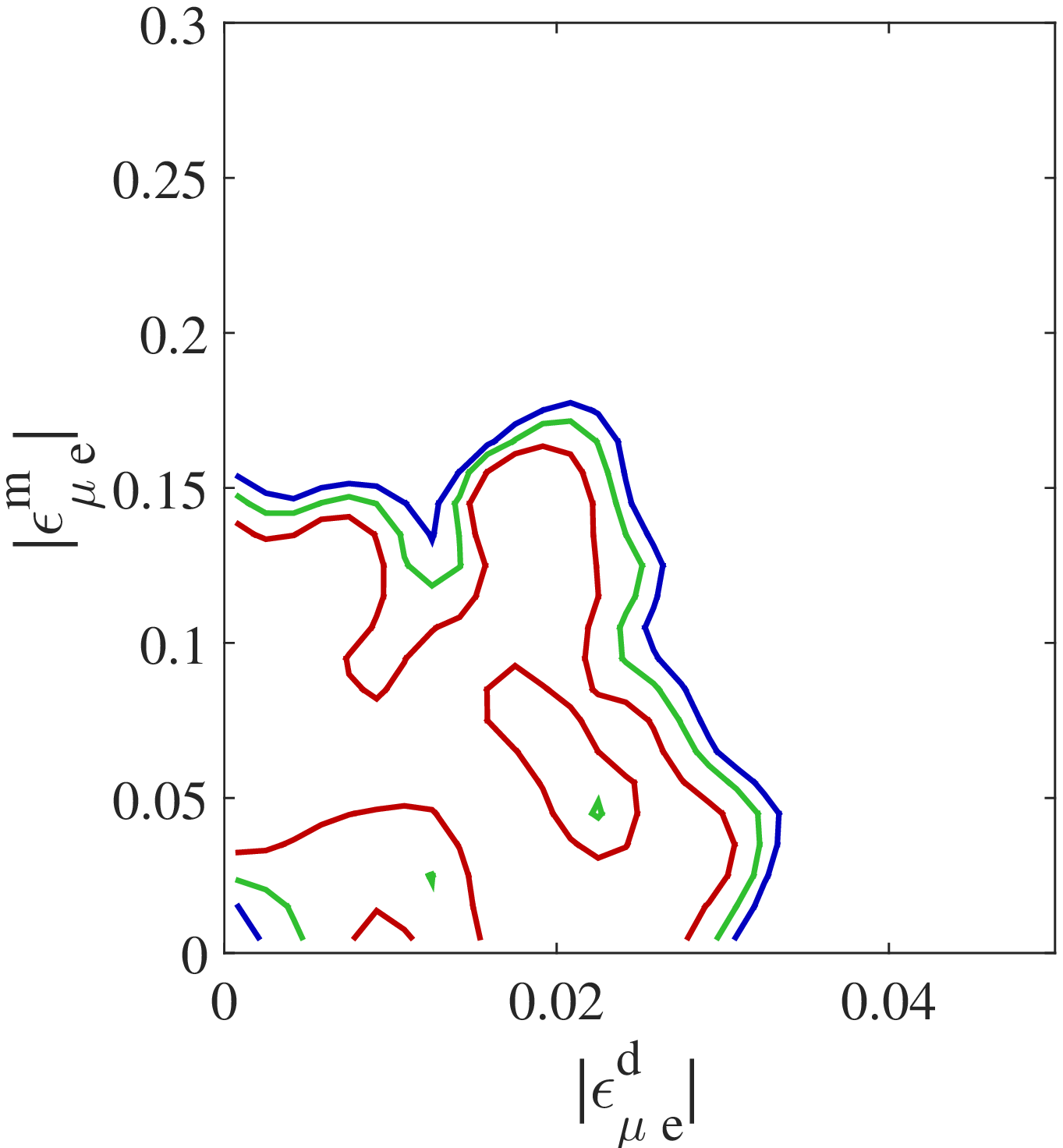, height=0.18\textheight, width=0.33\textwidth, bbllx=0,
bblly=0, bburx=480, bbury=440,clip=} &
\epsfig{file=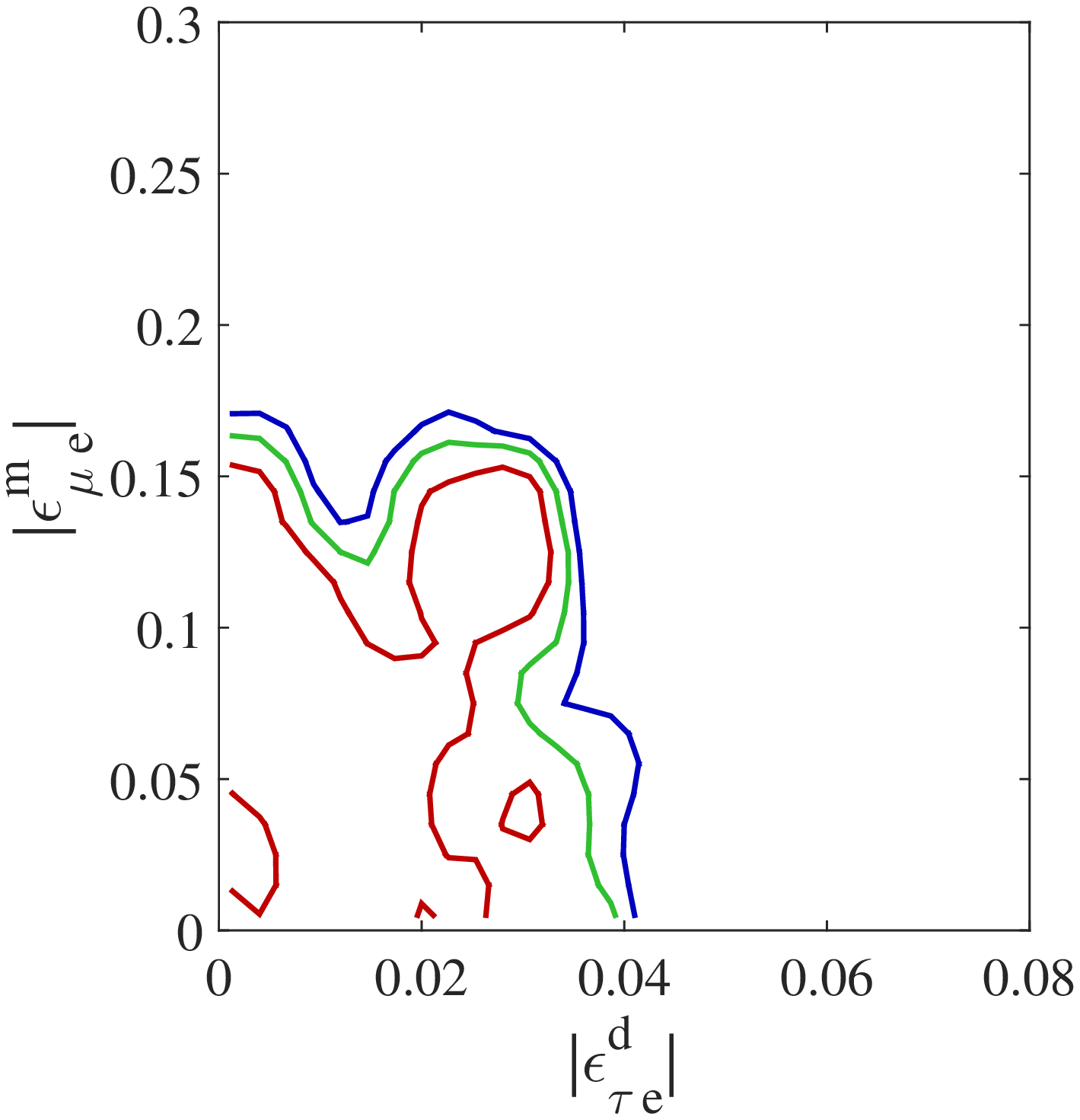, height=0.18\textheight, width=0.33\textwidth, bbllx=0,
bblly=0, bburx=480, bbury=440,clip=} \\ 
\epsfig{file=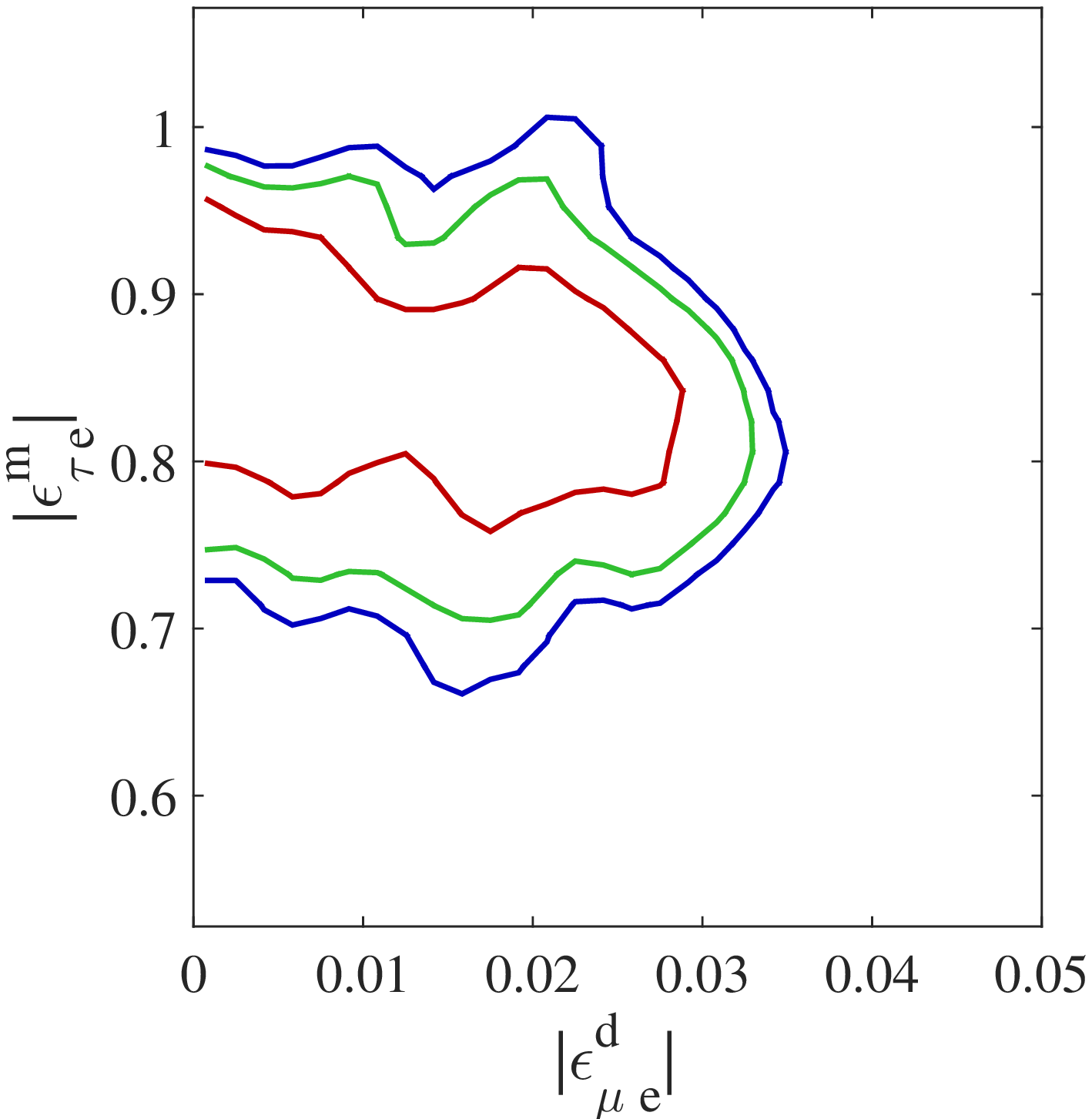, height=0.18\textheight, width=0.33\textwidth, bbllx=0,
bblly=0, bburx=480, bbury=440,clip=} &
\epsfig{file=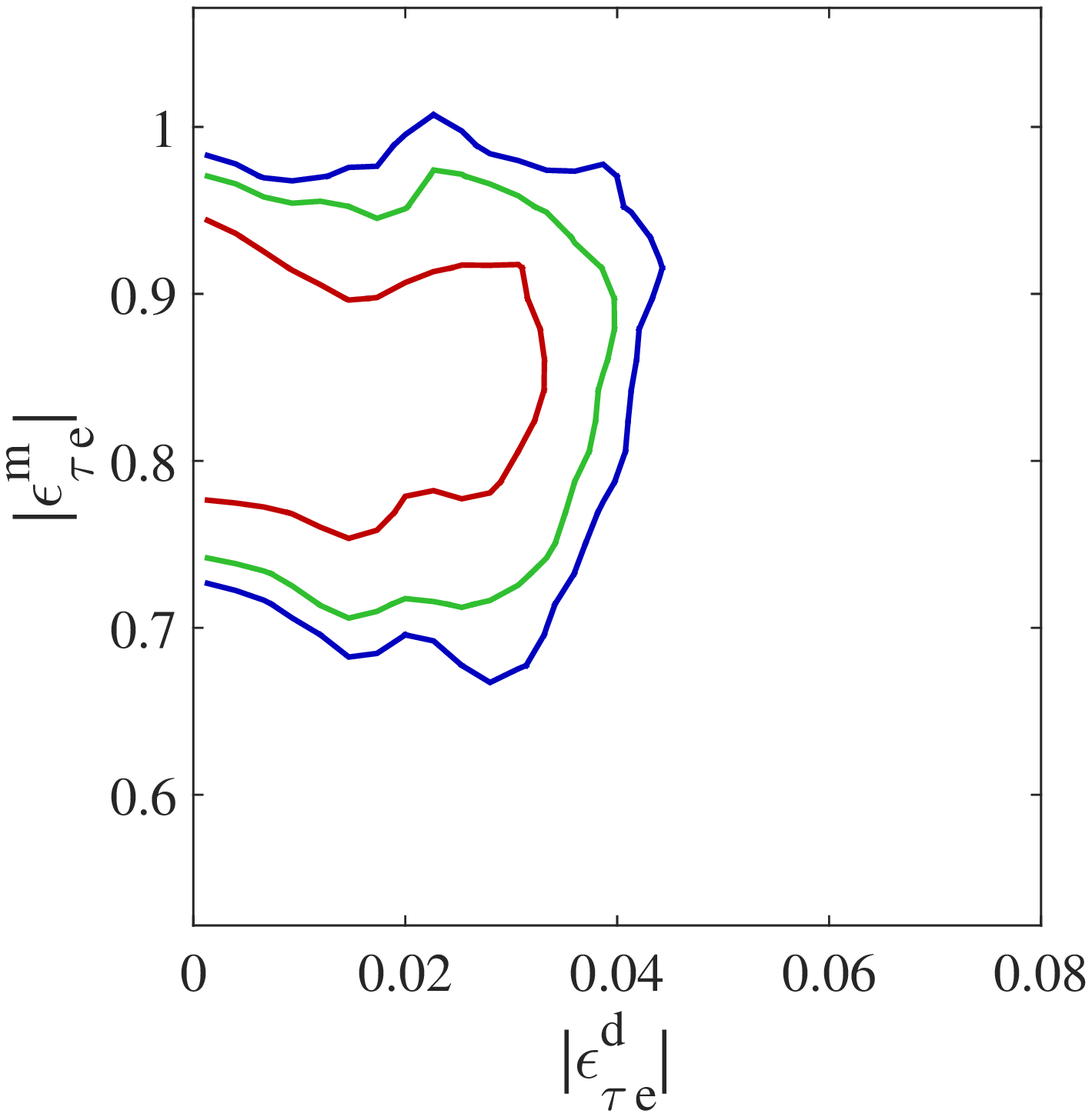, height=0.18\textheight, width=0.33\textwidth, bbllx=0,
bblly=0, bburx=480, bbury=440,clip=} \\ 
\epsfig{file=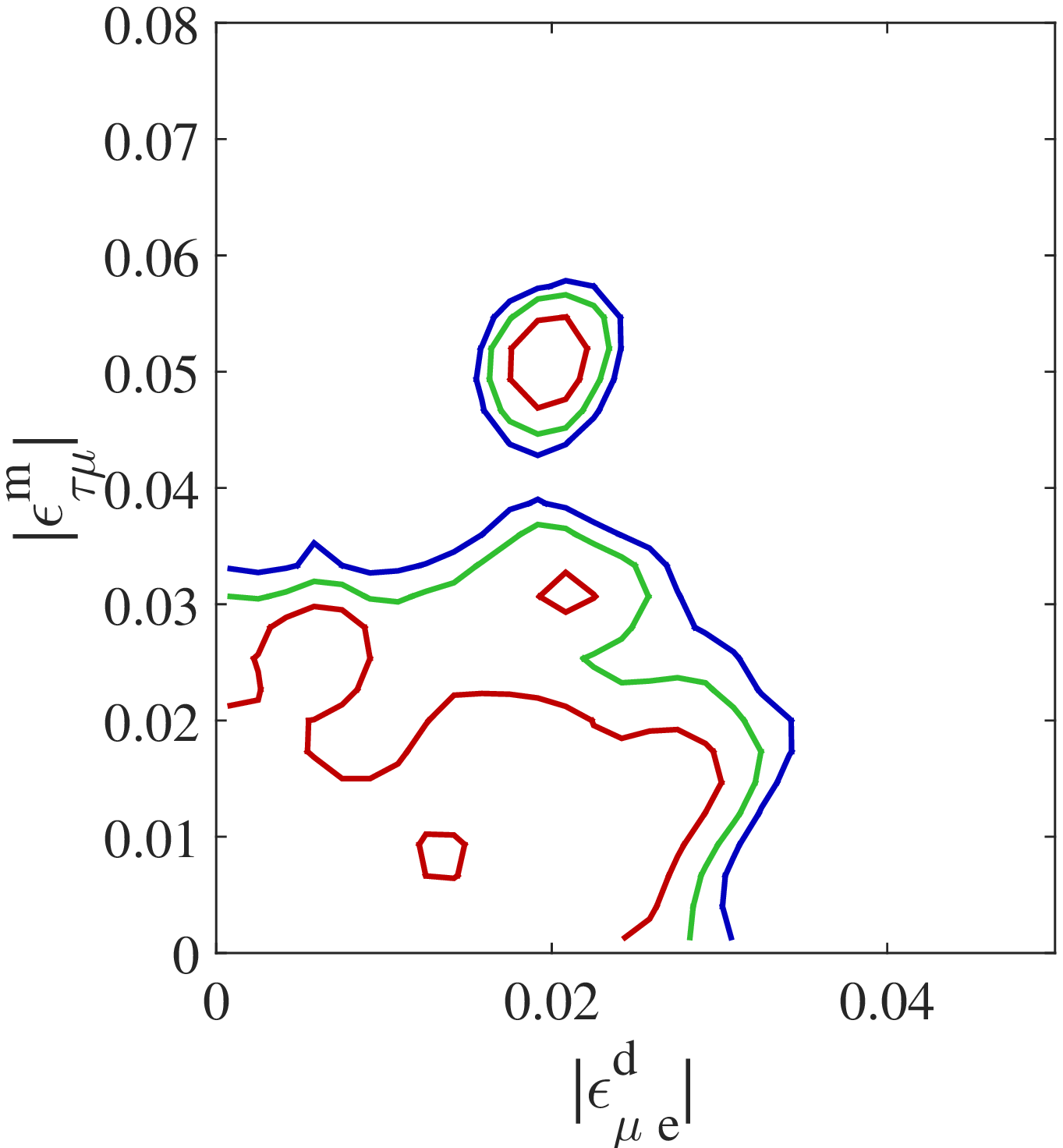, height=0.18\textheight, width=0.33\textwidth, bbllx=0,
bblly=0, bburx=480, bbury=440,clip=} &
\epsfig{file=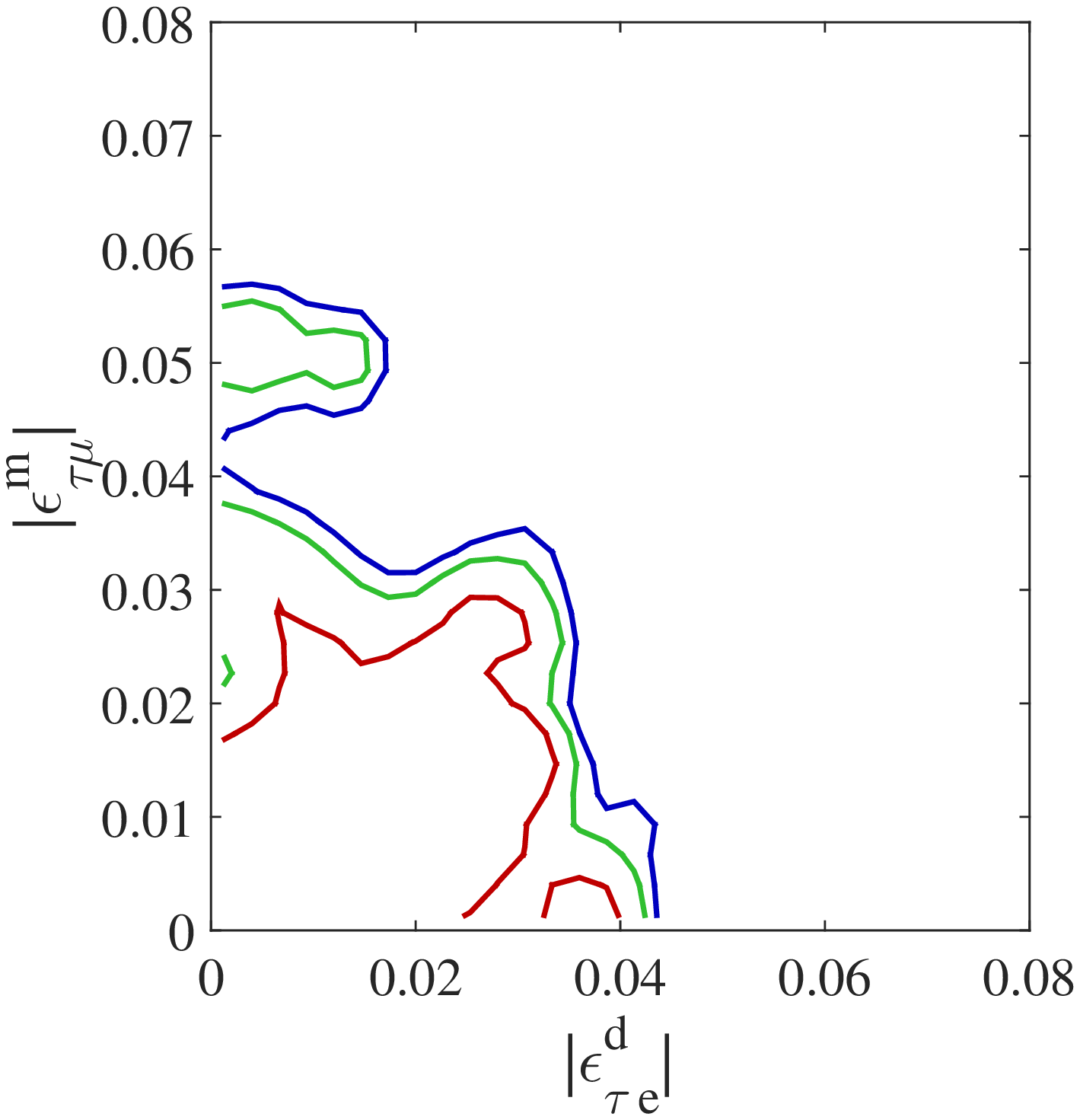, height=0.18\textheight, width=0.33\textwidth, bbllx=0,
bblly=0, bburx=480, bbury=440,clip=} \\ 
\epsfig{file=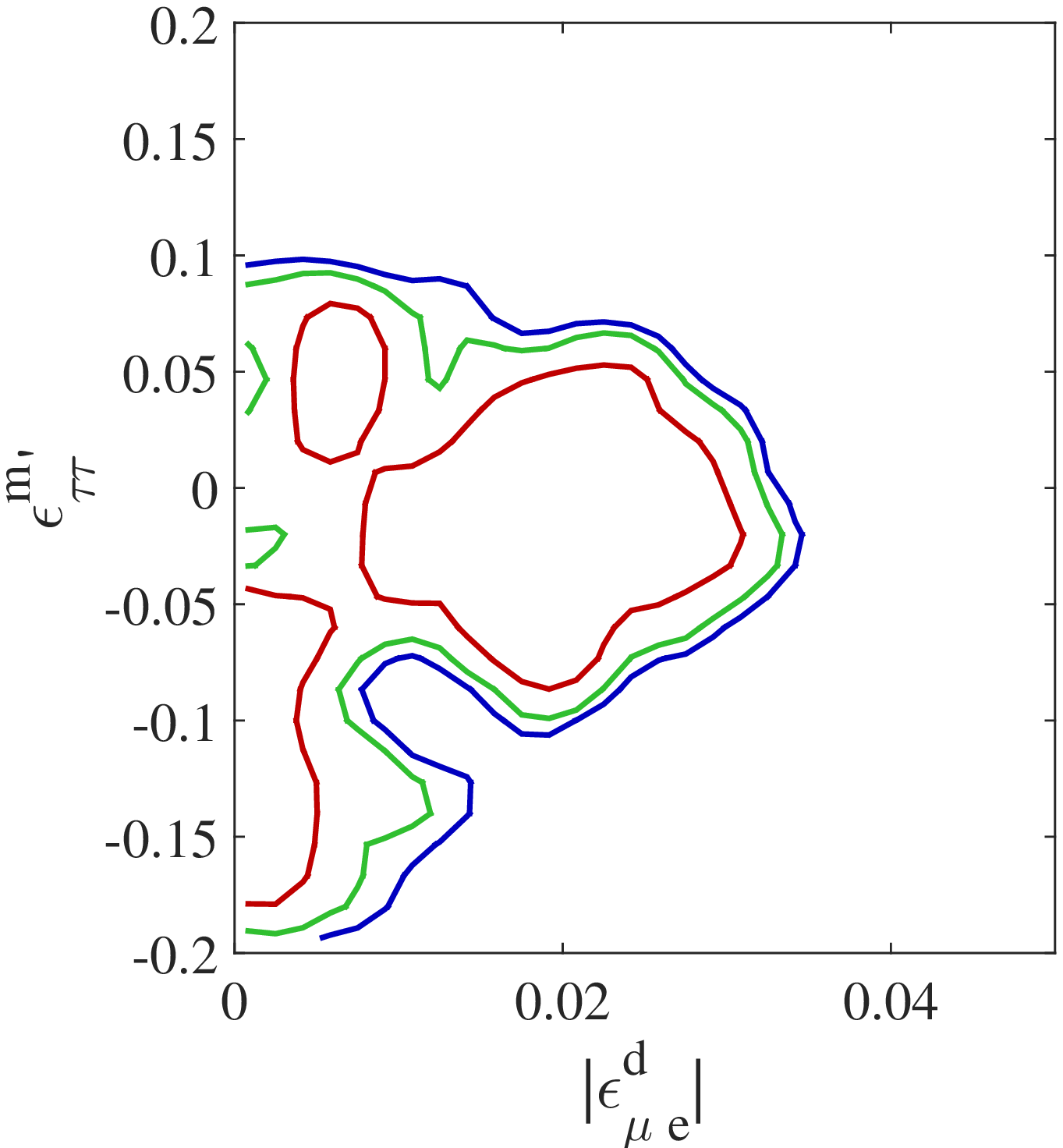, height=0.18\textheight, width=0.33\textwidth, bbllx=0,
bblly=0, bburx=480, bbury=440,clip=} &
\epsfig{file=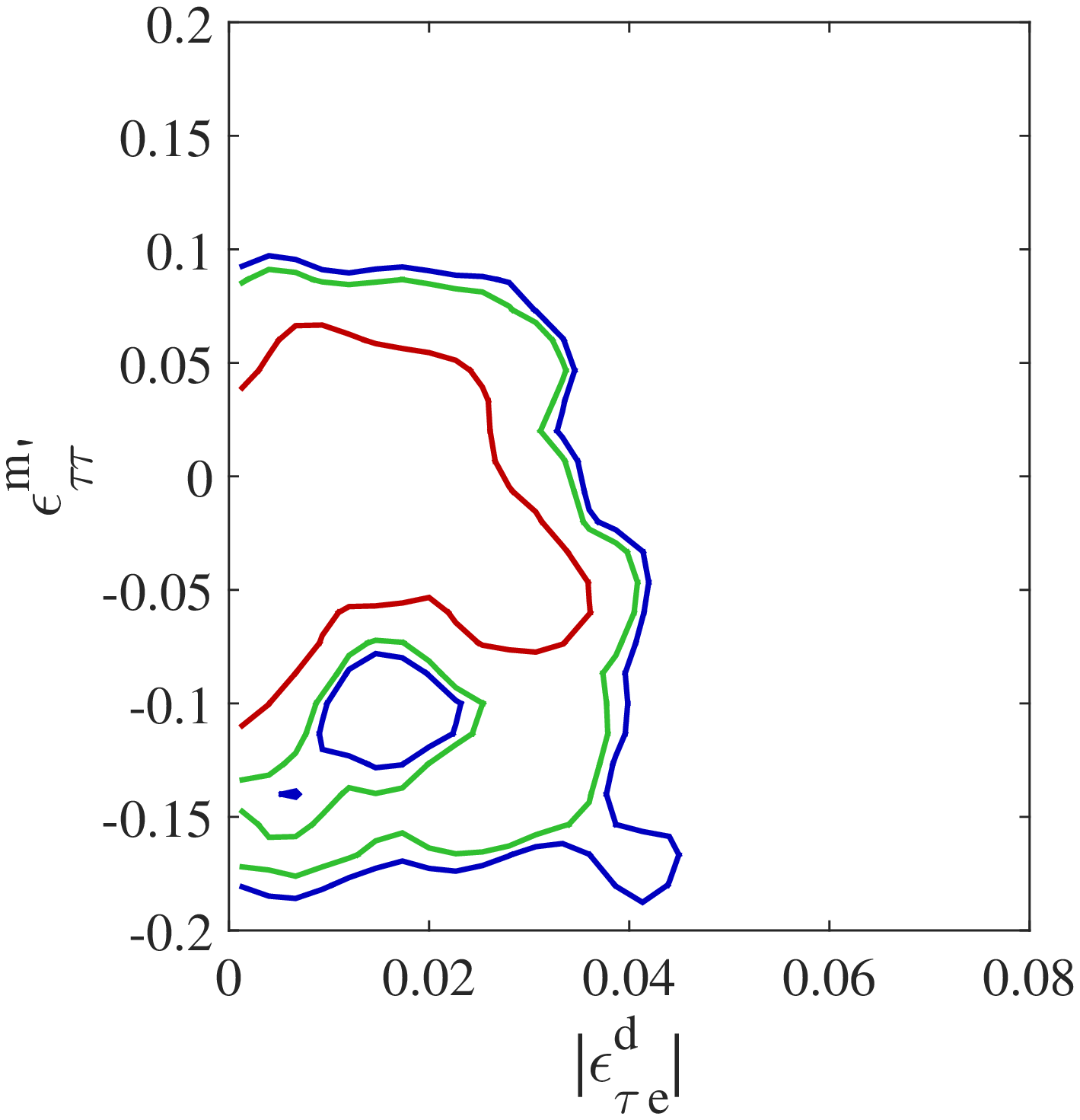, height=0.18\textheight, width=0.33\textwidth, bbllx=0,
bblly=0, bburx=480, bbury=440,clip=} 
\end{tabular}
\caption{\footnotesize Correlations between matter NSI parameters and detector 
NSI parameters with current bounds at DUNE. The 68~\% (red), 90~\% (green) and
95~\% (blue) credible regions are shown.}
\label{fig:allcorrelsd}
\end{figure}

Beyond the SM, CC-like and NC-like NSIs presumably arise from the same 
model of new physics. Therefore, it is natural to assume that both
source/detector 
and matter NSIs exist. It is interesting to probe the presence of 
correlations between various kinds of NSI parameters in neutrino oscillations. 
It is straightforward to pinpoint such correlations from the analytical
expressions 
for the oscillation probabilities given in Ref.~\cite{Kopp:2007ne}. The
non-standard terms 
in $\pmue$ up to linear order in $\sin\theta_{13}$ arising from $\epsdte$ and
$\epsmte$ are 
\begin{eqnarray}
  \pmue & \supset &
  - 4 \epsdte \tilde{s}_{13} s_{23}^{2} c_{23}
      \cos\dcp
      \left[   \sin^{2} \frac{AL}{4E}
             - \sin^{2} \frac{\dm{31} L}{4E}
             - \sin^{2} \frac{(\dm{31} - A)L}{4E}
      \right]                                           \nonumber\\
  && - 2 \epsdte \tilde{s}_{13} s_{23}^{2} c_{23}
      \sin\dcp
      \left[   \sin \frac{AL}{2E}
             - \sin \frac{\dm{31} L}{2E}
             + \sin \frac{(\dm{31} - A)L}{2E}
      \right]                                           \nonumber\\
  && + 4 \epsmte \tilde{s}_{13} s_{23}^{2} c_{23}
      \cos\dcp
      \left[   \sin^{2} \frac{A L}{4E}
             - \sin^{2} \frac{\dm{31} L}{4E}
             + \sin^{2} \frac{(\dm{31} - A)L}{4E}
      \right]                                           \nonumber\\
  && + 2 \epsmte \tilde{s}_{13} s_{23}^{2} c_{23}
      \sin\dcp
      \left[   \sin \frac{A L}{2E} 
             - \sin \frac{\dm{31} L}{2E}
             + \sin \frac{(\dm{31} - A)L}{2E}
      \right]                                           \nonumber\\
  && + 8 \epsmte \tilde{s}_{13} s_{23}^{2} c_{23}
      \cos\dcp \frac{A}{\dm{31} - A}
      \sin^{2} \frac{(\dm{31} - A)L}{4E}~, 
  \label{eq:correl1}
\end{eqnarray}
where we have assumed the NSI parameters to be real 
and used the notation $\tilde{s}_{13} = s_{13} \dm{31}/(\dm{31}-A)$. 
Close to the oscillation maximum, these terms can be combined into 
one term proportional to $\epsmte-\epsdte$.
Similarly, the terms involving $\epsdte$ and $\epsmme$ enter the formula as
\begin{eqnarray}
  \pmue & \supset &
  - 4 \epsdte \tilde{s}_{13} s_{23}^{2} c_{23}
      \cos\dcp
      \left[   \sin^{2} \frac{AL}{4E}
             - \sin^{2} \frac{\dm{31} L}{4E}
             - \sin^{2} \frac{(\dm{31} - A)L}{4E}
      \right]                                           \nonumber\\
  && - 2 \epsdte \tilde{s}_{13} s_{23}^{2} c_{23}
      \sin\dcp
      \left[   \sin \frac{AL}{2E}
             - \sin \frac{\dm{31} L}{2E}
             + \sin \frac{(\dm{31} - A)L}{2E}
      \right]                                           \nonumber\\
  && - 4 \epsmme \tilde{s}_{13} s_{23} c_{23}^{2}
      \cos\dcp
      \left[   \sin^{2} \frac{A L}{4E}
             - \sin^{2} \frac{\dm{31} L}{4E}
             + \sin^{2} \frac{(\dm{31} - A)L}{4E}
      \right]                                           \nonumber\\
  && - 2 \epsmme \tilde{s}_{13} s_{23} c_{23}^{2}
      \sin\dcp
      \left[   \sin \frac{A L}{2E} 
             - \sin \frac{\dm{31} L}{2E}
             + \sin \frac{(\dm{31} - A)L}{2E}
      \right]                                           \nonumber\\
  && + 8 \epsmme \tilde{s}_{13} s_{23}^{3}
      \cos\dcp \frac{A}{\dm{31} - A}
      \sin^{2} \frac{(\dm{31} - A)L}{4E} ~. 
  \label{eq:correl2}
\end{eqnarray}
These can be combined to give a term proportional to $\epsmme+\epsdte$ 
close to the oscillation maximum.
Thus, we expect to have a correlation between $\epsdte$ and $\epsmte$ and an 
anticorrelation between $\epsdte$ and $\epsmme$. The current bounds on the 
source/detector NSI parameters are more stringent than those on the matter NSI 
parameters. Therefore, in scanning over the parameter space, these correlations 
get washed out. However, if we make the assumption that these two types of NSI 
parameters have comparable bounds, then the correlations are visible. This can
be observed 
in the panels of Fig.~\ref{fig:correl}. In generating these plots, we have 
made use of the assumptions listed above, and used the (more stringent) priors of the source/detector 
NSI parameters for the matter NSI parameters as well. The true values assigned to the NSI parameters are half
of 
the bounds used. 
The correlations appear very weak because (a) the parameter space that has been 
scanned over is very large, (b) the conditions for the correlation include a 
very small value of $\theta_{13}$ and (c) the signal events have a spread in
energy 
away from the oscillation maximum. 
With the current bounds, which are very large for the matter NSI parameters, one
does 
not see any clear correlation between the two types of NSI parameters. In 
Figs.~\ref{fig:allcorrelss} and~\ref{fig:allcorrelsd}, we show the correlations
between all the source/detector 
and matter NSI parameters given their existing bounds. 

\subsection{Effect of statistics}

\begin{figure}
\epsfig{file=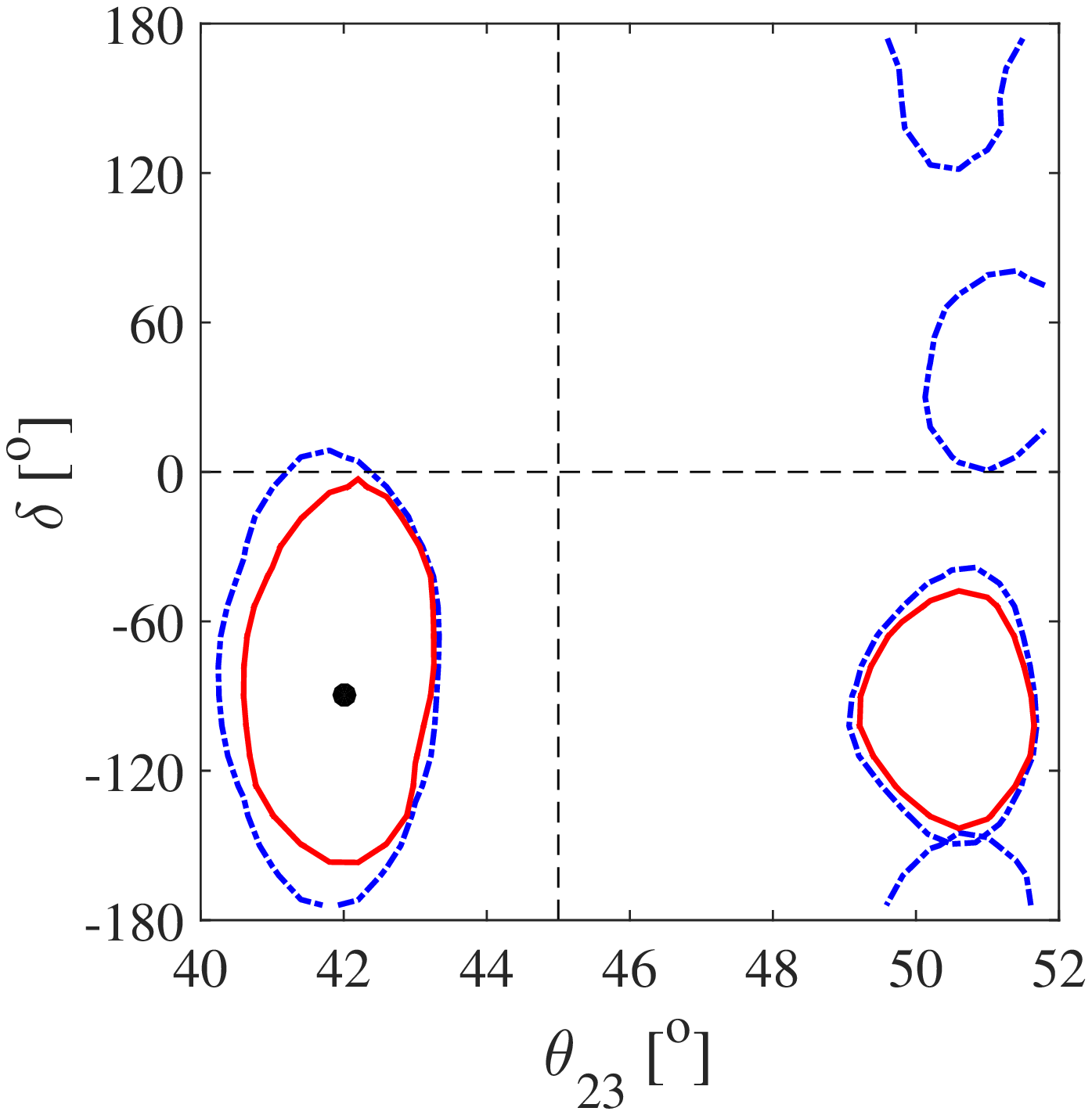, width=0.5\textwidth, bbllx=0, bblly=0,
bburx=436, bbury=414,clip=}
\caption{\footnotesize Sensitivity of DUNE in the $\theta_{23}-\dcp$ plane 
with 3+3 years (blue, dashed contours) and 5+5 years (red, solid contours) of
data. The simulated 
true values of these parameters are $42^\circ$ and $-90^\circ$, respectively. 
The contours enclose the 90~\% credible regions.}
\label{fig:stats}
\end{figure}

Finally, we make a note of the effect of statistics on the precision
measurements 
at DUNE in the presence of NSIs and on the measurement of these NSI parameters. 
In the preceding sections, we have assumed a runtime for DUNE of five years with 
neutrinos and antineutrinos each (5+5). In this section, we discuss the
results with lower 
statistics. In particular, we have re-run our simulations assuming three years
of 
running in each mode (3+3). Comparing results from the two cases
gives an indication of the 
effect of statistics on the results. 

In Fig.~\ref{fig:stats}, we have shown this comparison for the precision
measurements 
of $\theta_{23}$ and $\dcp$ at DUNE in the presence of all NSIs. The values of
the 
NSI parameters are taken to be the same non-zero values used in generating the
right 
panel of Fig.~\ref{fig:effect}. Comparing the contours corresponding to the cases of 5+5 and
3+3, we find that the precision in $\dcp$ is significantly worsened. It has
been shown that 
the combination of neutrino and antineutrino run helps in lifting the
$\theta_{23}-\dcp$ 
degeneracy, hence allowing for an accurate measurement of 
$\dcp$~\cite{Agarwalla:2013ju,Coloma:2014kca,Ghosh:2014zea,Ghosh:2015ena,Nath:2015kjg}. Usually,
it is 
sufficient to have a short run in the antineutrino mode that is just adequate
to 
resolve the degeneracy. In this case, however, the presence of NSIs introduces 
additional sources of CP violation. Therefore, there are additional regions of 
the parameter space that are allowed owing to the lower statistics. 

Furthermore, we have also checked the bounds imposed on the NSI parameters in
the case of 3+3. Once again, the experiment is seen to have much better
sensitivity 
to the matter NSI parameters than to the source/detector NSI parameters. 
We find that the bounds are only slightly worse than those 
shown in Table~\ref{tab:bounds} expected from the case of 5+5. Thus, the main 
advantage of collecting more data with DUNE is to determine the value of $\dcp$ 
with higher precision.

\section{Summary and conclusions}
\label{sec:sc}

The Deep Underground Neutrino Experiment (DUNE) is being proposed as a high precision next-generation neutrino experiment
to be built in the USA. The main physics goals of DUNE are to measure the neutrino
mass ordering, the octant of $\theta_{23}$ and the Dirac CP-violating phase $\dcp$. While the
baseline design for DUNE is known to be good for these physics goals
in the case of standard oscillations, it is worthwhile to recheck the
sensitivities of DUNE in the presence of non-standard neutrino interactions (NSIs). The NSIs are of two
kinds -- the charged-current-like source/detector NSIs and the neutral-current-like
matter NSIs. The source/detector NSIs affect the neutrino fluxes at the
production and detection of the neutrinos. On the
other hand, the matter NSIs play a role in the coherent scattering of the
neutrinos off the ambient matter during neutrino propagation. 
The role of the matter NSIs on the physics reach of DUNE
has been studied. However, it is more likely that if NSIs were to
exist, they 
would be both charged- as well as neutral-current-like. Therefore, in this
work, we have considered both source/detector and matter NSIs and looked at the
expected sensitivity of DUNE to neutrino oscillation parameters
as well as NSI parameters. We have performed this analysis by doing a full scan of the
entire relevant oscillation and NSI parameter space, 
which was accomplished by using a Markov Chain Monte Carlo. 
To the best of our knowledge, a complete phenomenological study of DUNE in the
presence of both source/detector and matter NSIs has not been performed before.

Through probability plots we have showed the impact of the NSI parameters on the
neutrino oscillation probability $P_{\mu e}$. Next, we have presented credible regions
in the $\theta_{23}-\dcp$ plane for the cases where we have taken in the fit (a) no NSIs, (b) only
source/detector NSIs, (c) only matter NSIs and (d) both source/detector and
matter NSIs together. The analysis was first performed for the case where 
the data were generated without the presence of NSIs. This was then repeated for the case
where NSIs were present in the generation of the data. In both cases, the
effect of source/detector NSIs on the $\theta_{23}$ and $\dcp$ measurements is
seen to be marginal, mainly because the source/detector NSIs are already severely
constrained by existing data and we have imposed priors in our analysis
corresponding to the existing constraints. The effect of matter NSIs is seen to
be larger, since the existing constraints on matter NSIs are weaker. Our
results, where we have included only matter NSIs, are observed to generally agree with those
obtained by 
others in the recent past, modulo the runtime assumed. The sensitivity to both
$\theta_{23}$ and $\dcp$ worsens when matter NSIs are included.
The worsening is more severe when the data are generated with NSIs compared to
when they are not. Finally, we have presented our results when
source/detector and matter NSIs are taken together. For the case where the data are generated 
with no NSIs, the credible regions with the inclusion of all NSIs in the fit are not 
very different from those with only matter NSIs. However, when the data are generated with 
NSIs, we observe a significant worsening of the precision with
the appearance of a fake 90~\% credible region in the $\theta_{23}-\dcp$ plane.
We have presented the 68~\%, 90~\% and 95~\% credible regions in the neutrino
oscillation parameter space in the presence of NSIs. The correlations between the
NSI parameters have been showcased. 

We have presented the sensitivities to the NSI parameters expected from DUNE. 
The 90~\% credible regions for the source/detector NSIs that we
expect are not much better than what we already know from current data. 
However, the expected sensitivities to the matter
NSI parameters are substantially stronger than the existing
bounds. 

We have presented our main results for a total of ten years of running of DUNE
-- five years in the neutrino mode and five years in the antineutrino mode (5+5). 
In order to study the impact of statistics, we have also presented the expected 
credible regions in the $\theta_{23}-\dcp$
plane for a DUNE run plan of three years in the neutrino mode and three years in the antineutrino
mode (3+3). Some of the earlier studies on DUNE with a full parameter scan have been done
for the case of 3+3~\cite{deGouvea:2015ndi,Coloma:2015kiu}. We have found that
for this case, many more degenerate solutions exist in the
$\theta_{23}-\dcp$ plane. This agrees with the results previously obtained in Ref.~\cite{Coloma:2015kiu} 
for the matter NSI only case. However, when the statistics are increased
to 5+5, the degenerate solutions for the matter NSI only
case disappear. 

To conclude, we have performed a complete analysis of the physics reach of DUNE
in the presence of both source/detector and matter NSIs, using a
full scan of the entire relevant parameter space. 
DUNE could improve the existing bounds on the matter NSIs, but would not be
able to improve the bounds on the source/detector NSIs. Increasing the
statistics from 3+3 to 5+5 reduces the negative impact of the matter NSIs on
the $\theta_{23}$ and $\dcp$ sensitivities. However, the correlation
between the source/detector and matter NSIs result in new degeneracies in the
$\theta_{23}-\dcp$ plane which remain even in the case of 5+5. 

\acknowledgments

This work was supported by the G{\"o}ran Gustafsson Foundation (M.B.). 
S.C.~acknowledges support from the Neutrino Project under the XII plan of 
Harish-Chandra Research Institute. 
This project has received funding from the European Union's Horizon 2020
research and innovation programme under the Marie Sklodowska-Curie grant
agreement No.~674896.

\bibliographystyle{apsrev}
\bibliography{dune_nsi}

\end{document}